\documentclass[11pt]{article}
\pdfoutput=1

\usepackage{amsmath}
\usepackage{amssymb}
\usepackage{amsfonts}
\usepackage{mathrsfs}

\usepackage{mathrsfs}
\usepackage{fullpage}
\usepackage{setspace}
\usepackage{bm}
\usepackage{bbm}
\usepackage{relsize}
\usepackage{adjustbox}
\usepackage{multirow}
\usepackage{cite}
\usepackage{filecontents}

\usepackage[normalem]{ulem}
\usepackage{enumerate}
\usepackage{yfonts}

\usepackage{psfrag}
\usepackage{array}
\usepackage{booktabs}
\newcolumntype{N}{>{\centering\arraybackslash}m{.5in}}
\newcolumntype{G}{>{\centering\arraybackslash}m{2in}}

\usepackage[latin1]{inputenc}
\usepackage{graphicx}
\usepackage{cancel}
\usepackage{slashed}
\usepackage{mathtools}

\usepackage[font=small,labelfont=bf]{caption}
\usepackage{subcaption}
\def\tg{\widetilde{g}}
\usepackage[colorlinks=true]{hyperref} 
\hypersetup{
    unicode=false,          
    pdftoolbar=true,        
    pdfmenubar=true,        
    pdffitwindow=false,     
    pdfstartview={FitH},    
    pdftitle={Gravitational wave from extreme mass-ratio inspirals as a probe
of extra dimensions},    
    pdfauthor={Mostafizur Rahman, Shailesh Kumar and Arpan Bhattacharyya},     
    pdfnewwindow=true,      
    colorlinks=true,       
    linkcolor=blue,          
    citecolor=red,        
    filecolor=magenta,      
    urlcolor=blue           
    }
\def\hJ{\hat{J_{z}}}
\def\hE{\hat{E}}
\def\hr{\hat{r}}

\def\bm{\overline{m}}

\def\Schld{Schwarzschild}

\def\RN{Reissner-Nordstr\"{o}m}

\usepackage{hyperref}
\hypersetup{colorlinks=true}

\def\equationautorefname~#1\null{%
	Eq.~(#1)\null
}
\def\figureautorefname~#1\null{%
	Fig.~#1\null
}
\def\tableautorefname~#1\null{%
	Table.~#1\null
}
\def\sectionautorefname~#1\null{%
	Section #1\null
}
\def\appendixautorefname~#1\null{%
	Appendix #1\null
}

\begin{document}

\numberwithin{equation}{section}
{
\begin{titlepage}
\begin{center}

\hfill \\
\hfill \\
\vskip 0.75in

{\Large \bf Gravitational wave from extreme mass-ratio inspirals as a probe of extra dimensions
}\\

\vskip 0.3in

{\large
Mostafizur Rahman${}$\footnote{\href{mailto: mostafizur.r@iitgn.ac.in}{mostafizur.r@iitgn.ac.in}}, Shailesh Kumar${}$\footnote{\href{mailto: shailesh.k@iitgn.ac.in}{shailesh.k@iitgn.ac.in}} and Arpan Bhattacharyya${}$\footnote{\href{mailto: abhattacharyya@iitgn.ac.in}{abhattacharyya@iitgn.ac.in}}}

\vskip 0.3in

{\it ${}$Indian Institute of Technology, Gandhinagar, Gujarat-382355, India}

\vskip.5mm

\end{center}

\vskip 0.35in

\begin{center} 
{\bf ABSTRACT }
\end{center}
The field of gravitational waves is rapidly progressing due to the noticeable advancements in the sensitivity of gravitational-wave detectors that has enabled the detection prospects of binary black hole mergers. Extreme mass-ratio inspiral (EMRI) is one of the most compelling and captivating binary systems in this direction, with the detection possibility by the future space-based gravitational wave detector. In this article, we consider an EMRI system where the primary or the central object is a spherically symmetric static braneworld black hole that carries a \textit{tidal charge} $Q$. We estimate the effect of the tidal charge on total gravitational wave flux and orbital phase due to a non-spinning secondary inspiralling the primary. We further highlight the observational implications of the tidal charge in EMRI waveforms. We show that LISA (Laser Interferometer Space Antenna) observations can put a much stronger constraint on this parameter than black hole shadow and ground-based gravitational wave observations, which can potentially probe the existence of extra dimensions. 
\vfill


\end{titlepage}
}

\newpage

\section{Introduction}
The observational facets of gravitational waves (GWs), upholding Einstein's fundamental prediction set almost a century ago, have prompted researchers to inspect the diverse aspects of black hole spacetimes \cite{PhysRevLett.116.061102, Carson:2020rea, Berti:2015itd, Perkins:2020tra, Barack:2018yly, LIGOScientific:2021sio}. It has enabled scientists to investigate numerous characteristics of the strong regime of massive compact sources by opening up new avenues for GW astronomy. It also provides a first-hand to test the existing theories of gravity. Cataclysmic events such as binary black hole mergers, neutron star collisions, and supernova explosions are the most common and prominent sources of GWs. In this direction, \textit{extreme mass-ratio inspirals} (EMRIs), consisting of a stellar-mass object (secondary) orbits a massive black hole (primary), have recently gained considerable attention for analyzing the gravitational wave signal to accurately test the predictions of general relativity in the strong regime of gravity \cite{Amaro-Seoane:2007osp, Gair:2017ynp, Babak:2017tow,Gair:2010yu, PhysRevD.78.064028}. The secondary is being treated as a background perturbation to the spacetime regulated by the primary, where we maintain the tiny mass-ratio ($q \equiv \mu/M = 10^{-7}-10^{-4}$) for the setup; $\mu$ is the mass of the stellar mass object, and $M$ is the mass of the primary massive black hole. GW signal generated through EMRIs provides a chance to measure various fascinating properties of supermassive black holes \cite{PhysRevD.69.082005, Arun_2009}. In order to examine the characteristics of a supermassive black hole, EMRI systems have been proven one of the most suitable testing grounds in this direction \cite{PhysRevD.52.5707, PhysRevD.69.124022, Glampedakis_2006, PhysRevD.75.042003, PhysRevD.77.024035, PhysRevD.81.024030}. Current developments focusing on such studies imply a possible detection of these sources through space-borne detectors like  Laser Interferometer Space Antenna Experiments (LISA) \cite{Amaro-Seoane:2007osp, Babak:2017tow, Amaro_Seoane_2012, Luo_2016}. 
The detection prospects of EMRIs with LISA can yield accurate measurements of various features of the binaries by probing the fundamental physics \cite{Barack_2019, Barausse2020}, even if parameter estimation encounters a few challenging issues \cite{PhysRevD.101.044027, PhysRevD.95.103012}. 
Furthermore, since the secondary remains in the strong gravity region of the supermassive object for a large number of wave cycles ($\mathcal{O}(10^4-10^5)$) before merger \cite{PhysRevD.78.064028}, the emitted gravitational waveform from such a system is expected to contain very accurate information about the geometry surrounding the supermassive black hole \cite{Babak:2017tow, PhysRevD.52.5707, PhysRevD.75.042003, Maselli:2020zgv}. Thus EMRIs are ideally suited to test general relativity (GR), and constraint different modified theories of gravity \cite{PhysRevD.75.042003,Gair:2012nm, Yunes:2011aa, Canizares:2012ji}. In this paper, we are considering the effect of higher dimensions on the EMRI waveform and checking whether LISA can probe their existence.

The idea that there may be more than three spatial dimensions in our universe is almost as old as GR. In the 1920s, Kaluza and Klein proposed the existence of compact extra dimensions in an attempt to unify gravity with electromagnetism \cite{Kaluza:1921tu, Klein:1926tv}. After that, string theory also provides a compelling justification for considering space as multidimensional \cite{polchinski_1998, Horava:1995qa}. The method by which these extra dimensions are concealed is a key issue in multidimensional theories. In this direction, the concept of braneworld has become very prominent \cite{Rubakov:2001kp, Antoniadis:1998ig,Randall:1999ee, Randall:1999vf, Maartens2004}. According to this model, we live on a four-dimensional slice (called the brane) of a higher dimensional $Z_2$-symmetric bulk spacetime. All the matter fields are confined to this brane except for gravity which can propagate in all extra dimensions. These extra dimensions can be large \cite{Antoniadis:1998ig} or even infinite, as illustrated in Randall-Sundrum model \cite{Randall:1999ee, Randall:1999vf}, where a single brane is embedded in a five-dimensional anti-de Sitter spacetime. In the low energy limit, the effective field equations on the brane resemble the Einstein field equations with two correction terms. One is a local correction term originating from the $Z_2$ symmetry of the spacetime and Israel junction conditions, while the other is a non-local correction term originating from bulk Weyl tensor \cite{Shiromizu:1999wj}.
Interestingly, the static, spherically symmetric vacuum solutions of the effective field equations have the mathematical form of \RN\ black holes in general relativity \cite{Dadhich:2000am, Chamblin:2000ra}. However, these black holes carry tidal charges, which can take negative values, unlike the electric charge of \RN\ black holes \footnote{We like to point out that it is difficult to obtain black hole solutions in the brane that does not exhibit pathological behaviour in the bulk \cite{PhysRevLett.87.231302, PhysRevD.63.064015, PhysRevD.65.084010, PhysRevD.61.065007}. However, in this paper, we will work in such a regime so that we can effectively neglect the effect of the bulk.}. The observational consequences of tidal-charged black holes have been discussed in the literature in the context of black hole shadow and gravitational wave observations \cite{Chakravarti:2019aup, Banerjee:2019nnj, Horvath:2012ru, Zakharov:2018awx, Banerjee:2019sae, Neves:2020doc, Chakraborty:2021gdf, Mishra:2021waw, Visinelli:2017bny, Vagnozzi:2019apd, Vagnozzi:2022moj, Khlopunov:2022ubp, Khlopunov:2022jaw}. In the present context, we consider the effect of tidal charge on the EMRI waveform and check whether LISA observation can put a tighter constraint on this parameter and, consequently, on the extra dimensions.
 
Let us briefly get into the overview of the article. In Section~(\ref{BH}), we provide the spacetime metric carrying a tidal charge with proper motivation and study the equatorial motion and innermost stable circular orbit (ISCO) of a non-spinning secondary object moving in the vicinity of a primary black hole. In Section~(\ref{pert1}), we discuss Teukolsky perturbation equations and source terms. This sets up the stage for estimating the GW fluxes and computing the adiabatic evolution of the orbital dynamics of the secondary provided in Section~(\ref{teu1}), Section~(\ref{apenteu3}) and Section~(\ref{teu2n}). In the next Section (\ref{rslt}), we numerically analyze the effect of tidal charge on our results, including the gravitational waveform and estimate the mismatch with respect to the Schwarzschild limit. In the last Section~(\ref{dscn}), we discuss the implications of our study and future prospects. We also provide Appendix~ (\ref{apenteu1}) and Appendix~(\ref{t2n}) for computational details. 
\par 
\textit{Notation and Convention: } We set the fundamental constants $G$ and $c$ to unity and  adopt positive sign convention $(-1,1,1,1)$. Upper-case Roman letters are used to denote higher dimensional indices, and Greek letters are used to represent four-dimensional indices.
\section{Brane black hole with a tidal charge and orbital motion}\label{BH}

In this section, we provide the metric of the spacetime describing the primary and discuss the equatorial orbital motion of the secondary object. Here, we adopt the geometrical approach put forward by Shiromizu, Maeda, and Sasaki to describe the effective field equation in the brane \cite{Shiromizu:1999wj}. In this approach, the five-dimensional bulk spacetime is governed by the Einstein field equations. The brane is the fixed point of the reflection symmetry of the bulk spacetime. The induced metric on the brane takes the following form $g_{AB}=\tg_{AB}-n_A n_B$, where $\tg_{AB}$ is the metric of the bulk spacetime and $n_{A}$ is the spacelike unit normal to the brane. Owing to the reflection symmetry of the bulk spacetime and invoking the Gauss-Cadazzi relation, we can write the effective field equation on the brane as \cite{Shiromizu:1999wj, Maartens:2001jx}
\begin{equation}\label{EFE}
   G_{\mu\nu}+E_{\mu\nu}=8\pi \left[T_{\mu\nu}+\frac{6}{\lambda_b}\Pi_{\mu\nu}\right]
\end{equation}
where $G_{\mu\nu}$ is the Einstein tensor on the brane, $E_{\mu\nu}$ is electric part of the bulk Weyl tensor,
 $T_{\mu\nu}$ is the energy-momentum tensor, $\lambda_b$ is the brane tension, and $\Pi_{\mu\nu}$ is the ``squared'' energy-momentum tensor which contains terms like, $T_{\mu}^{\alpha}T_{\nu\alpha}$, $TT_{\mu\nu}$ etc. As can be seen, the consideration of higher dimensional spacetime accompanies two non-trivial correction terms to the standard Einstein field equation: $\Pi_{\mu\nu}$, which contains information about the local bulk effects on the matter on the brane and a traceless tensor $E_{\mu\nu}$, which is originated from bulk Weyl tensor and carries information about the gravitational field outside the brane \cite{Shiromizu:1999wj}. Note that, in the limit $\lambda_b\to\infty$, $E_{\mu\nu}$ and $\frac{6}{\lambda_b}\Pi_{\mu\nu}$ vanish, and we recover the standard general relativity. The vacuum, static, and spherically symmetric solution of the above equation can be written in the following form \cite{Dadhich:2000am},
\begin{align}\label{m1}
    ds^{2} = -f(r) dt^{2} + \frac{1}{f(r)} dr^{2} + r^{2} (d\theta^{2}+\sin^{2}\theta d\phi^{2}),
\end{align}
where 
\begin{equation}
    f(r)\equiv \dfrac{\Delta}{r^2} = \frac{r^2-2Mr+\beta}{r^2}
\end{equation}
where $M$ is the mass of the object and $\beta$ is some constant of integration. Note that the above metric is identical to a \RN\ black hole if we take $\beta=e^2$ with $e$ being the electric charge. However, there is a certain distinction; we are considering a vacuum spacetime, i.e., there is no Maxwell field outside the black hole. The black hole spacetime instead carries a ``tidal'' charge, which contains information about the bulk Weyl tensor. Furthermore, this tidal charge can have both positive and negative values. Intuitively, this is because the tidal charge parameter originates from the bulk Weyl tensor; and thus depends on the mass on the brane. Hence, the tidal charge strengthens the brane's gravitational field, unlike the electric charge that weakens it \cite{Maartens:2001jx}. Interestingly, for the negative values of the tidal charge parameter $\beta=-QM^2$ ($Q$ is positive definite), there is only one positive solution $r_+=M+M\sqrt{1+Q}$ to $f(r)=0$, corresponding to the event horizon of the black hole. This is in contrast to \RN\ black holes, which have a Cauchy horizon inside their event horizon. Since the Cauchy horizon marks the boundary of the domain of dependence, tidally charged black hole spacetime with $\beta=-QM^2$ does not have the pathological features of \RN\ black holes \cite{Dafermos:2003wr, PhysRevLett.67.789, Cardoso:2017soq, Rahman:2018oso}. Furthermore, these black holes have interesting astrophysical implications \cite{Chakravarti:2019aup, Banerjee:2019nnj, Banerjee:2019sae, Banerjee:2021aln}. Motivated by these facts, we consider only negative values of the tidal charge and its effect on the EMRI dynamics. Henceforth, we denote $\beta=-QM^2$ to emphasize this point.

We consider the primary object is described by such braneworld black holes and study the equatorial, quasi-circular motion of the secondary object moving in its vicinity. For the sake of simplicity, we model the secondary object as a point particle with the energy-momentum tensor \cite{Poisson:2011nh}
\begin{equation}\label{SET}
    T^{\mu\nu}=\mu \int d\tau \frac{\delta^{(4)}\left(x-z(\tau)\right)}{\sqrt{-g}} u^\mu u^{\nu},
\end{equation}
where $\mu$ is mass of the object, $\tau$ is the proper time along the worldline $z^{\mu}$, and $u^{\mu}=dz^{\mu}/d\tau$ represents the tangent to the line. We aim to examine the orbital dynamics, which gives us the innermost circular stable orbit (ISCO) that further helps us to investigate the evolution of the secondary in the EMRI system.

Without loss of generality, we take the orbital motion on the equatorial plane defined by $\theta = \frac{\pi}{2}$. One can further express an equation of the following form,
\begin{align}
    -f(r)\dot{t}^{2}+\frac{1}{f(r)}\dot{r}^{2}+r^{2}\dot{\phi}^{2} = -1, \label{norme1}
\end{align}
where we denote the derivative with respect to proper time ($\tau$). The spacetime under consideration exhibits two Killing vectors ($\frac{\partial}{\partial t}, \frac{\partial}{\partial \phi}$) that have two constants of motion ($E, J$) called as energy and angular momentum of the object. Also, it is useful to introduce dimensionless quantities $\hat{E} = \frac{E}{\mu}$, $\hat{J}=\frac{J}{M\mu}$ and $\hat{r}=\frac{r}{M}$, which will be convenient for computational purpose.
\begin{align}
    \hat{E} = f(\hat{r}) \hat{\dot{t}} \hspace{7mm} ; \hspace{7mm} \hat{J} = \hr^{2}{\dot{\phi}}. \label{const1}
\end{align}
Using Eq.(\ref{const1}), the Eq.(\ref{norme1}) can be organized as,
\begin{align}
    \hat{\dot{r}}^{2}+\hat{V}_{eff} (\hat{r}) = \hat{E}^{2}, 
\end{align}
where
\begin{align}
    \hat{V}_{eff} (\hat{r}) = \left(1+\frac{\hat{J}^2}{\hat{r}^2}\right) f(\hat{r}).
\end{align}
The effective potential $\hat{V}_{eff}(\hat{r})$ determines the orbital motion of the object. For stable, circular orbits with radius $\hat{r}_{0}$, we note that $\hat{V}_{eff}(\hat{r}_{0})=0$ and $\frac{d\hat{V}_{eff}}{d\hat{r}}\vert_{\hat{r}=\hat{r}_{0}}=0$; where we examine the stability through $\frac{d^{2}\hat{V}_{eff}}{d\hat{r}^{2}}\vert_{\hat{r}=\hat{r}_{0}}<0$. This determines the value of $\hat{E}$ and $\hat{J}$ for stable circular orbits. We obtain the innermost stable circular orbit (ISCO) by imposing $\frac{d^{2}\hat{V}_{eff}}{d\hat{r}^{2}}\vert_{\hat{r}=\hat{r}_{0}}=0$.
Further, in order to estimate gravitational wave fluxes, we also require orbital frequency $\hat{\Omega}\equiv M\Omega=d\phi/d\hat{t}$ of a circular equatorial orbit 
\begin{align}
    \hat{\Omega} = \frac{\hat{J}}{\hat{r}^{2}\hat{E}}f(\hat{r}).
\end{align}
Here, one needs to set values of $\hat{E}$ and $\hat{J}$. For the computational purpose, we use the numerical values of ISCO radius ($\hat{r}^{ISCO}$) and orbital frequency $\hat{\Omega}$ that help us in calculating the gravitational wave flux. We shall use the related quantities at a later stage.

\section{Gravitational wave fluxes and perturbation equations}\label{pert1}

The evolution of the secondary object in the vicinity of the primary and generated gravitational wave fluxes can be studied through perturbation methods. In this section, we elaborate on gravitational wave fluxes produced by an EMRI system. We point out that we shall be considering a small mass-ratio for all the subsequent analyses, i.e., $q\equiv \mu/M\ll 1$. Under this assumption, we now determine the evolution of the secondary through the perturbation method. In this setup, the secondary object perturbs the background spacetime  $ g_{\mu\nu}^{(0)}$ governed by the line element presented in Eq.~(\ref{m1}) as $g_{\mu\nu}=g_{\mu\nu}^{(0)}+q h_{\mu\nu}$, where $h_{\mu\nu}$ is the perturbative correction term. The perturbed field equation can then be written as follows
\begin{equation}
     \delta G_{\mu\nu}+\delta E_{\mu\nu}=8\pi \left[T_{\mu\nu}+\frac{6}{\lambda_b}\Pi_{\mu\nu}\right]
\end{equation}
where $T_{\mu\nu}$ is governed by Eq.~(\ref{SET}). Note that $\Pi_{\mu\nu}$ is squared energy-momentum tensor, and thus quadratic in $q$. Hence, in the linear order of $q$, we can neglect the contribution of this term. Furthermore, we consider that perturbation on the brane has a negligible effect on the projected Weyl tensor, i.e., $\delta E_{\mu\nu}=0$. This can be justified by taking a low energy limit. We assume that the matter-energy density on the brane is much smaller than the brane tension \cite{Kanno:2002ia}. Under this assumption, the perturbed Weyl tensor becomes $\delta E_{\mu\nu}\approx (l/L) G_{\mu\nu}\ll G_{\mu\nu}$, where $l$ is the curvature length scale of the bulk spacetime and $L$ is the curvature length scale of the supermassive black hole on the brane \cite{Kanno:2003au, Kanno:2003sc}. Hence, we can neglect the contribution of the term. The perturbation causes the secondary object to follow a forced geodesic equation, and as a result, the system enters the inspiral phase. In this paper, we analyze the problem within the framework of adiabatic approximation \cite{PhysRevD.78.064028, PhysRevD.103.104014}.

The motivation further comes from the fact that the orbital time scale is substantially shorter than the dissipative time scale. This allows us to approximate the orbits of the secondary as geodesics over a brief period of time where the timescale is much less than the inspiral timescale. Further, the time-averaged, dissipative portion of the self-force controls how quickly the orbit's energy and angular momentum vary \cite{ PhysRevD.103.104014},
\begin{equation}\label{adiabatic}
	\begin{aligned}
		\left(\frac{d\hat{E}}{dt}\right)^{\textrm{orbit}}=-\Bigg\langle\frac{d\hat{E}}{dt}\Bigg\rangle_{\textrm{GW}} \hspace{3mm} ; \hspace{3mm}  	\left(\frac{d\hat{J}}{dt}\right)^{\textrm{orbit}}=-\Bigg\langle\frac{d\hat{J}}{dt}\Bigg\rangle_{\textrm{GW}},
	\end{aligned}
\end{equation}
where $<\hspace{1mm}>$ depicts the time-averaging over a period that is substantially longer than the time evolution of the orbital parameters but smaller than the inspiral time scales. 
The Teukolsky equation is solved to determine the back-reaction impact on the orbit that provides the adiabatic evolution of the secondary object. Here, we only analyze the adiabatic motion. However, the adiabatic approximation breaks down once the object passes the ISCO and transits to a geodesic plunge orbit \cite{PhysRevD.62.124022}. Let us now look at the details of the Teukolsky perturbation equation and how one can compute gravitational wave fluxes.

\subsection{Teukolsky Equation}\label{teu1}
In this section, we briefly describe the Teukolsky formalism \cite{PhysRevLett.29.1114, 1972ApJ, 1973ApJ, Teukolsky:1974yv} which we shall use in this paper. The formalism is most suitable for Petrov type D black holes where only the non-vanishing Weyl scalar is $\Psi_{2}$ \cite{Pound:2021qin, PhysRevD.103.124057}. The major advantage of the Teukolsky equation is that it is a scalar equation that describes the dynamics of the Weyl scalars ($\Psi_{4}, \Psi_{0}$)- tetrad projections of the Weyl curvature tensor, where $\Psi_{4}$ and $\Psi_{0}$ provide the outgoing and ingoing radiation respectively at the asymptotic region. For our interest, we would be investigating the effects for $\Psi_{4}$ as GW detectors measure outgoing waves at asymptotic infinity. At a large distance from the source, the information about the two polarizations of GWs is captured in $\Psi_{4}$,
\begin{align}
    \Psi_{4} = \frac{1}{2}\partial^{2}_{\hat{t}}(h_{+}-ih_{\times}),
\end{align}
where $h_{+}$ and $h_{\times}$ are two GW polarizations that encode the information of GW strain, $\partial^{2}_{\hat{t}}$ denotes the double partial derivative with respect to time. Further, the Teukolsky equation in terms of various NP quantities and different operators are written in the following form \cite{PhysRevLett.29.1114, Degollado2010, PhysRevD.104.104006},
\begin{align}\label{teu2}
[ (\mathbf{\Delta}-(3\gamma-\gamma^{*}+4\mu+\mu^{*}))(D+\rho - 4\epsilon) - (\delta^{*}-(\beta^{*}+3\alpha+4\pi-\tau^{*}))(\delta-4\beta+\tau) + 3\Psi_{2}] \Psi_{4} = -4\pi T_{4},
\end{align}
together with the source term $T_{4}$,
\begin{align}\label{source1}
  T_{4} =& (\mathbf{\Delta}-(3\gamma-\gamma^{*}+4\mu+\mu^{*}))[(\delta^{*}+2\tau^{*}-2\alpha)T_{nm^{*}}-(\mathbf{\Delta}-(2\gamma-2\gamma^{*}+\mu^{*}))T_{m^{*}m^{*}}]+ \nonumber \\ 
  & (\delta^{*}-(\beta^{*}+3\alpha+4\pi-\tau^{*}))[(\mathbf{\Delta}-2\gamma-2\mu^{*})T_{nm^{*}}-(\delta^{*}-(2\beta^{*}+2\alpha-\tau^{*}))T_{nn} ],
\end{align}
where various quantities ($\gamma, \gamma^{*}, \mu, \mu^{*}, \epsilon, \rho, \tau, \tau^{*} \beta, \beta^{*}, \alpha, \pi$) and operators $\mathbf{\Delta}=n^{\mu}\partial_{\mu}$, $D=l^{\mu}\partial_{\mu}$, $\delta=m^{\mu}\partial_{\mu}$, $\delta^{*}=\bar{m}^{\mu}\partial_{\mu}$ depend on NP tetrad; related expressions of the same for the metric under consideration are given in appendix~(\ref{apenteu1}). Once the Eq.~(\ref{teu2}) is separated into radial and angular parts, the solution for $\Psi_{4}$ in the Fourier space can be written as
\begin{align}\label{Psi_4}
\Psi_4=\rho^4\sum_{l=2}^{\infty}\sum_{m=-l}^{l}\int_{-\infty}^{\infty}d\hat{\omega} R_{lm\hat{\omega}}(\hat{r})~{}_{-2}S_{lm\hat{\omega}}(\theta) e^{i(m\phi-\hat{\omega} \hat{t})}
\end{align}
where ${}_{-2}S_{lm\hat{\omega}}$ is the spin weighted spheroidal harmonics with weight $-2$ satisfying the angular Teukolsky equation
\begin{align}\label{ang_Teuk}
\Bigg[&\frac{1}{\sin\theta}\frac{d}{d\theta}\left(\sin\theta\frac{d}{d\theta}\right)-\left(\frac{m-2\cos\theta}{\sin\theta}\right)^2-2+\lambda_{lm\hat{\omega}}\Bigg]{}_{-2}S_{lm\hat{\omega}}=0,
\end{align}
where $\lambda_{lm\hat{\omega}}$ is the  separation constant.  The eigenfunction of the angular Teukolsky equation complies with the normalization requirement: 
$\int \sin\theta d\theta d\phi |S_{lm\hat{\omega}} e^{im\phi}|^2=1$. Here, we have omitted suffix $-2$ in ${}_{-2}S_{lm\hat{\omega}}$ for the sake of brevity. Note that, in spherically symmetric spacetimes, the spin-weighted spheroidal harmonics $S_{lm\hat{\omega}}$ and the separation constant $\lambda_{lm\hat{\omega}}$ are independent of $\hat{\omega}$. However, to have a notational consistency with \cite{PhysRevD.102.024041}, we continue to write them as $S_{lm\hat{\omega}}$ and $\lambda_{lm\hat{\omega}}$.  The radial equation is given as,
\begin{align} \label{thomo}
    \hat{\Delta}^{2}\frac{d}{d\hat{r}}\Big(\frac{1}{\hat{\Delta}}\frac{dR_{lm\hat{\omega}}}{d\hat{r}}\Big)-V(\hat{r})R_{lm\hat{\omega}} = \mathcal{T}_{lm\hat{\omega}},
\end{align}
with
\begin{align}\label{pot1}
    V(\hat{r}) = -\frac{K^{2}+4i(\hat{r}-1)K}{\hat{\Delta}}+8i\hat{\omega} \hat{r}+\lambda_{lm\hat{\omega}} \hspace{5mm} ; \hspace{5mm} K = \hat{r}^{2}\hat{\omega},
\end{align}
where the source term $\mathcal{T}_{lm\hat{\omega}}$ is given in the appendix (\ref{apenteu2}). In order to compute $\lambda_{lm\hat{\omega}}$ and spin-weighted spheroidal harmonics, we use \texttt{Black Hole Perturbation Toolkit package} \cite{BHPT}. The homogeneous radial equation, mentioned in Eq.~(\ref{pot1}), provides two linearly independent solutions namely $R^{in}_{lm\hat{\omega}}$ and $R^{up}_{lm\hat{\omega}}$ which take the asymptotic values at the horizon ($r_{+}$) and at the infinity as, 
\begin{equation}\label{rty1}
  R^{in}_{lm\hat{\omega}}(\hat{r})\sim\begin{cases}
    B^{tran}_{lm\hat{\omega}}\hat{\Delta}^{2}e^{-i\hat{\omega} \hat{r}_{*}}; & \text{$\hat{r}\longrightarrow \hat{r}_{+}$},\\
    B^{out}_{lm\hat{\omega}}\hat{r}^{3}e^{i\hat{\omega} \hat{r}_{*}}+B^{in}_{lm\hat{\omega}}\hat{r}^{-1}e^{-i\hat{\omega} \hat{r}_{*}}; & \text{$\hat{r}\longrightarrow \infty$},
  \end{cases}
\end{equation}
\begin{equation}\label{try2}
  R^{up}_{lm\hat{\omega}}(\hat{r})\sim\begin{cases}
    D^{out}_{lm\hat{\omega}}e^{i\hat{\omega} \hat{r}_{*}}+D^{in}_{lm\hat{\omega}}\hat{\Delta}^{2}e^{-i\hat{\omega} \hat{r}_{*}}, & \text{$\hat{r}\longrightarrow \hat{r}_{+}$},\\
    D^{tran}_{lm\hat{\omega}}\hat{r}^{3}e^{i\hat{\omega} \hat{r}_{*}}; & \text{$\hat{r}\longrightarrow \infty$}.
  \end{cases}
\end{equation}
where $\frac{d\hat{r}_{*}}{d\hat{r}}=\frac{\hr^{2}}{\hat{{\Delta}}}$ with $\hat{\Delta}=\hat{r}^{2}f(\hat{r})$ and $\hr_{\pm}=1\pm\sqrt{1+Q}$. We find the $R^{in}_{lm\hat{\omega}}$ and $R^{up}_{lm\hat{\omega}}$ using Sasaki-Nakamura formalism \cite{SASAKI198185, SASAKI198268, 10.1143/PTP.67.1788} described in  Section~($\ref{apenteu3}$). With the homogeneous solutions in our hand, we can use the Green function method to solve the radial Teukolsky equation; for which the solution takes the following form,
\begin{align}
    R_{lm\hat{\omega}}(\hat{r}) = \frac{1}{W}\Bigg(R^{up}_{lm\hat{\omega}}\int_{\hat{r}_{+}}^{\hat{r}}\frac{R^{in}_{lm\hat{\omega}}\mathcal{T}_{lm\hat{\omega}}}{\hat{\Delta}^{2}}d\hat{r}+R^{in}_{lm\hat{\omega}}\int_{\hat{r}}^{\infty}\frac{R^{up}_{lm\hat{\omega}}\mathcal{T}_{lm\hat{\omega}}}{\hat{\Delta}^{2}}d\hat{r}\Bigg).
\end{align}
The Wronskian is: $W=\Big(R^{in}_{lm\hat{\omega}}\partial_{\hat{r}_{*}}R^{up}_{lm\hat{\omega}}-R^{up}_{lm\hat{\omega}}\partial_{\hat{r}_{*}}R^{in}_{lm\hat{\omega}} \Big)$. Since the solution is purely ingoing at the horizon and outgoing at the infinity, the radial function behaves as

\begin{align}
    R^{in}_{lm\hat{\omega}}(\hat{r})=\begin{cases}
    \mathcal{Z}^{\infty}_{lm\hat{\omega}}\hat{\Delta}^{2}e^{-i\omega \hat{r}_{*}}; & \text{$\hat{r}\longrightarrow \hat{r}_{+}$},\\
    \mathcal{Z}^{H}_{lm\hat{\omega}}\hat{r}^{3}e^{i\omega \hat{r}_{*}}e^{i\hat{\omega} \hat{r}_{*}}; & \text{$\hat{r}\longrightarrow \infty$}, \end{cases}
\end{align}
where
\begin{align}\label{amp_def}
		\mathcal{Z}_{lm\hat{\omega}}^{H,\infty}=\mathcal{C}_{lm\hat{\omega}}^{H,\infty}\int_{\hat{r}_{+}}^{\infty
		}d\hat{r}\frac{R^{in,up}_{lm\hat{\omega}}\mathcal{T}_{lm\hat{\omega}}}{\Delta^{2}} \hspace{3mm} ; \hspace{3mm} C^{\infty}_{lm\hat{\omega}} = \frac{B^{tran}_{lm\hat{\omega}}}{2i\hat{\omega} B^{in}_{lm\hat{\omega}}D^{tran}_{lm\hat{\omega}}} \hspace{3mm} ; \hspace{3mm} C^{H}_{lm\hat{\omega}} = \frac{1}{2i\hat{\omega} B^{in}_{lm\hat{\omega}}}.
\end{align}
$\mathcal{Z}^{H,\infty}$ are amplitudes that finally gives gravitational wave flux. Other parameters like $B^{tran}_{lm\hat{\omega}}$ and $D^{tran}_{lm\hat{\omega}}$ are arbitrary and we determine them in the Section~(\ref{apenteu3}). Using the definition of the secondary's stress-energy tensor, the amplitudes are written as
\begin{equation}\label{amp_form}
	\begin{aligned}
		\mathcal{Z}^{H,\infty}_{l m \hat{\omega}}
		=\mathcal C^{H,\infty}_{l m \hat{\omega}}\int_{-\infty}^{\infty} d\hat{t}\, e^{i(\hat \omega \hat t-m\,\phi(t))}I^{H,\infty}_{l m \hat{\omega}}[\hat{r}(\hat{t}),\hat{\theta}(\hat{t})],
	\end{aligned}
\end{equation}
with
\begin{align}\label{amp_integrant}
	I^{H,\infty}_{lm\hat{\omega}}[\hat{r}(\hat{t}),\hat{\theta}(\hat{t})]
	=&\Big[A_0-A_1\frac{d}{d\hat{r}}+A_2\frac{d^2}{d\hat{r}^2}\Big]R^{\textrm{in},\textrm{up}}_{l m \hat{\omega}}\Big|_{\hat{r}(\hat{t}),\hat{\theta}(\hat{t})}.
\end{align}
The coefficients $A_0,~ A_1$ and $A_2$ have been presented in the Eq.~(\ref{app12}). Now using Eq.~(\ref{Psi_4}) and Eq.~(\ref{amp_form}), we can obtain gravitational wave signal,
\begin{align}\label{gw_signal}
		h=-\frac{2}{\hat{r}}\sum_{l=2}^{\infty}\sum_{m=-l}^{l}\int_{-\infty}^{\infty}\frac{d\hat{\omega}}{\hat{\omega}^2}~\mathcal Z^{H}_{lm\hat{\omega}}(\hat{r})~S_{lm\hat{\omega}}(\vartheta) e^{i(m\varphi-\hat{\omega}(\hat{t}-\hat{r}_*))}.
\end{align}
The angle between the axis of symmetry, $z$-axis, of the main object and the line of sight of a viewer at infinity, is denoted by the symbol $\vartheta$; also $\varphi \equiv \phi(\hat{t}=0)$. Since we are interested in examining the circular equatorial orbits that significantly simplify the source term. The amplitude Eq.~(\ref{amp_form}) also gets simplified with the consideration $\phi(\hat{t})=\hat{\omega} \hat{t}$, i.e., $\mathcal{Z}^{H,\infty}_{l m \hat{\omega}}
=\mathcal A^{H,\infty}_{l m \hat{\omega}}\delta(\hat{\omega}-m\hat{\Omega})$ at $\hat{r}_0$, with  $\mathcal A^{H,\infty}_{l m \hat{\omega}}=2\pi~\mathcal C^{H,\infty}_{l m \hat{\omega}} I^{H,\infty}_{l m \hat{\omega}}(\hat{r}_0,\pi/2)$. With this, the gravitational waveform takes the following form,
\begin{align}\label{signal_equatorial}
		h=-\frac{2}{\hat{r}}\sum_{l=2}^{\infty}\sum_{m=-l}^{l}\frac{\mathcal A^{H}_{lm\hat{\omega}}}{(m\hat{\Omega})^2}~S_{lm\hat{\omega}}(\vartheta) e^{im(\varphi-\hat{\Omega}(\hat{t}-\hat{r}_*))}.
\end{align}
We obtain energy flux from Eq.~(\ref{signal_equatorial}) that can be integrated over the solid angle. We also make use of the property of amplitude  $\mathcal{Z}_{l -m-\hat{\omega}}^{H,\infty}=(-1)^{l}\bar{\mathcal{Z}}_{lm\hat{\omega}}^{H,\infty}$ that controls the sum over $m$ in Eq.~(\ref{signal_equatorial}) to positive values of $m$. The energy flux at infinity becomes,

\begin{align}\label{flux_inf}
	\left(\frac{d\hat{E}}{d\hat{t}}\right)^{\infty}_{\textrm{GW}}=\sum_{l=2}^{\infty}\sum_{m=1}^{l}\frac{|\mathcal A^{H}_{lm\hat{\omega}}|^2}{2\pi( m\hat{\Omega})^2} \hspace{3mm} ; 
\end{align}
where sum over $m$ goes as $m=1, 2, ..., l$. In a similar way, the energy flux at the horizon is given by,
\begin{align}\label{flux_hor}
		\left(\frac{d\hat{E}}{d\hat{t}}\right)^{H}_{\textrm{GW}}=\sum_{l=2}^{\infty}\sum_{m=1}^{l}\alpha_{l m}\frac{|\mathcal A^{\infty}_{lm\hat{\omega}}|^2}{2\pi( m\hat{\Omega})^2} \hspace{3mm} ;
\end{align}
where, $\alpha_{lm}=\frac{1}{|C_{lm}|^2}\Big(256(2\hat{r}_{+})^{5}\hat{\kappa} (\hat{\kappa}+4\epsilon^{2})(\hat{\kappa}+16\epsilon^{2})(m\hat{\Omega})^{3}\Big)$ with $\hat{\kappa} = \hat{\omega}$, $\epsilon = \frac{1}{4\hat{r}_{+}}$ and $|C_{lm}|^{2}=\lambda_{lm\Omega}^{2}(\lambda_{lm\Omega+2})^{2}+144(m\hat{\Omega})^{2}$.
\subsection{Sassaki-Nakamura Equation}\label{apenteu3}

In this section, we elaborate on  Sasaki-Nakamura (SN) formalism \cite{SASAKI198185, SASAKI198268, 10.1143/PTP.67.1788} required for the computation of the energy flux as mentioned in Section~(\ref{teu1}) for the metric under consideration. The solutions of the homogeneous part of the radial equation mentioned in Eq.~(\ref{thomo}) diverge at infinity because of the long-range character of the potential. SN formalism helps us to find an appropriate transformation that makes the potential short-range, which is much more convenient from the perspective of numerical computation \cite{Kojima:1983ua}. The SN equation is written as,
\begin{equation}
\bigg[f(\hat{r})^2\frac{d^2}{d \hat{r}^2}+f(\hat{r}) \bigg(\frac{d{f(\hat{r})}}{d\hat{r}} - F(\hat{r}) \bigg) 
\frac{d}{d\hat{r}} -U(\hat{r})\bigg] 
X_{l m \hat{\omega}}=0 \ , \label{eq:SNeq}
\end{equation}
with $f(\hat{r})= \frac{d\hat{r}}{d\hat{r}^*}= \frac{\hat{\Delta}}{\hat{r}^2}$.
The coefficient $F(\hat{r})$ is defined as
\begin{equation}
F(\hat{r})= \frac{\eta(\hat{r})_{,\hat{r}}}{\eta(\hat{r})}\frac{\hat{\Delta}}{\hat{r}^2} \ ,
\end{equation}
where ${}_{,\hat{r}}$ denotes the derivative with respect to $\hat{r}$ and 
\begin{equation}
\eta(\hat{r}) = c_0 + \frac{c_1}{\hat{r}} + \frac{c_2}{\hat{r}^2} + \frac{c_3}{\hat{r}^3} + \frac{c_4}{\hat{r}^4} \ ,
\end{equation}
with
\begin{align}\label{c0}
c_0 = (\lambda_{lm\hat{\omega}}+2) \lambda_{lm\hat{\omega}}-12 i\hat{\omega} \hspace{1mm} ; \hspace{1mm} 
c_1 = -24i Q \hat{\omega}  \hspace{1mm} ; \hspace{1mm}
c_2 = -12 Q \hspace{1mm} ; \hspace{1mm}
c_3 = 24 Q \hspace{1mm} ; \hspace{1mm}
c_4 = 12Q^2 
\end{align}
The function $U(\hat{r})$ is,
\begin{equation}
U(\hat{r}) = \frac{\hat{\Delta} U_1(\hat{r})}{\hat{r}^4}+G(\hat{r})^2 + \frac{\hat{\Delta} 
G(\hat{r})_{,\hat{r}}}{\hat{r}^2}  - F(\hat{r})G(\hat{r}) \ ,
\end{equation}
together with,
\begin{align}
G(\hat{r}) &= - \frac{2(\hat{r}-1)}{\hat{r}^2} + \frac{\hat{\Delta}}{\hat{r}^3} \hspace{3mm} ; \hspace{3mm}
U_1(\hat{r}) = V(\hat{r}) +\frac{\hat{\Delta}^2}{\beta}\Big[\Big(2 \alpha + \frac{\beta_{,\hat{r}}}{\hat{\Delta}} 
\Big)_{\!\!,\hat{r}} -\frac{\eta(\hat{r})_{,\hat{r}}}{\eta(\hat{r})}\Big(\alpha + \frac{\beta_{,\hat{r}}}{\hat{\Delta}}\Big) 
\Big]  \ ,\\
\alpha &= -i K(\hat{r}) \frac{\beta}{\hat{\Delta}^2}  + 3i K(\hat{r})_{,\hat{r}} + \lambda_{\ell m \hat{\omega}} + 
\frac{6\hat{\Delta}}{\hat{r}^2} \hspace{3mm} ; \hspace{3mm} \beta =2 \hat{\Delta}\Big[ -i K(\hat{r}) + \hat{r} - 1 - \frac{2\hat{\Delta}}{\hat{r}} \Big]  \ .
\end{align}
$K(\hat{r})$ and $V(\hat{r})$ are given in Eq.~(\ref{pot1}). The SN equation has two linearly independent solutions, $X^{in}_{lm\hat{\omega}}$ and $X^{up}_{lm\hat{\omega}}$, asymptotic behaviour of them is given by
\begin{equation}
X^{in}_{lm\omega}
\sim
\begin{cases}
e^{-i \hat{\omega} \hat{r}^\ast} \quad  &\hat{r} \to \hat{r}_+  \\
A^{\textup{out}}_{l m \hat{\omega}}  e^{i \hat{\omega} \hat{r}^\ast} + A^{\textup{in}}_{l m 
\hat{\omega}} e^{-i \hat{\omega} \hat{r}^\ast} \quad 
&\hat{r} \to \infty  \label{eq:inBCSN}
\end{cases}\,,
\end{equation}
\begin{equation}
X^{up}_{lm\omega}\sim
\begin{cases}
 C^{\textup{out}}_{l m \hat{\omega}} e^{i \hat{\omega} \hat{r}^{\ast}} + C^{\textup{in}}_{l m 
\hat{\omega}}e^{- i \hat{\omega} \hat{r}^{\ast}} \quad \,  &r\to r_+   \label{eq:upBCSN} \\
e^{i \hat{\omega} \hat{r}^{\ast}}  \, \quad &\hat{r} \to \infty  
\end{cases}\,.
\end{equation}
The solutions of SN equations and Teukolsky are related by
\begin{equation}
\begin{aligned}
R^{\textup{in},\textup{up}}_{l m \hat{\omega}}(\hat{r}) =\frac{1}{\eta}\bigg[\bigg(\alpha 
+\frac{\beta_{,\hat{r}}}{\hat{\Delta}}\bigg)Y^{\textup{in},\textup{up}}_{l m \hat{\omega}} - 
\frac{\beta}{\hat{\Delta}}{Y^{\textup{in},\textup{up}}_{l m \hat{\omega}}}_{\!\!, \hat{r}}\bigg]  \ \hspace{3mm} ; \hspace{3mm}
Y^{\textup{in},\textup{up}}_{\ell m \hat{\omega}}& =\frac{\hat{\Delta}}{\hat{r}}  
X^{\textup{in},\textup{up}}_{l m \hat{\omega}} \ . \label{rinsol}
\end{aligned}
\end{equation}
Using the above relations, one can find the arbitrary parameters $D^{\textup{tran}}_{l m \hat{\omega}}$ and $B^{\textup{tran}}_{lm\omega} $: 
\begin{equation}
D^{\textup{tran}}_{l m \hat{\omega}}= - \frac{4\hat{\omega}^2}{c_0} \hspace{3mm} ; \hspace{3mm} B^{\textup{tran}}_{l m 
\hat{\omega}}  = \frac{1}{d_{l m\hat{\omega}}} .
\end{equation}
Here we avoid mentioning the expression for $d_{lm\omega}$ due to its long and complicated structure. We consider its numerical values by taking the distinct values of tidal charge $Q$. However, the idea of obtaining the arbitrary functions $D^{tran}_{lm\hat{\omega}}$ and $B^{tran}_{lm\hat{\omega}}$ comes from acquiring the relation between $R^{in, up}_{lm\hat{\omega}}$ and $X^{in, up}_{lm\hat{\omega}}$ given by Eq.(\ref{rinsol}). 
It is straightforward to calculate these functions except the $d_{lm\hat{\omega}}$. We focus on the approach as proposed in \cite{10.1093/ptep/ptaa149}. In order to compute the quantity $d_{lm\hat{\omega}}$, the purely incoming solution should have its higher order corrections with respect to ($r-r_{+}$). Therefore, the asymptotic form of the ingoing solution can be written up to $\mathcal{O} ((r-r_{+})^{2})$ in the following form,
\begin{align}
X^{in, H}_{lm\hat{\omega}} = \Big( 1+ a_{1} (r-r_{+})+a_{2} (r-r_{+})^{2} + \mathcal{O} ((r-r_{+})^{3}) \Big) e^{-i\omega r^{*}} \label{dlmasye1}, 
\end{align}
where $X^{in, H}_{lm\hat{\omega}}$ denotes the ingoing solution near the horizon ($H$). Also, coefficients $a_{1}$ and $a_{2}$ are established by satisfying the Eq.(\ref{eq:SNeq}). Once $a_{1}$ and $a_{2}$ are determined, we can obtain $B^{tran}_{lm\hat{\omega}}$ by putting Eq.(\ref{dlmasye1}) into Eq.(\ref{rinsol}). Finally,
using boundary conditions mentioned in Eq.~(\ref{eq:inBCSN}) and Eq.~(\ref{eq:upBCSN}) and upon integrating Eq.(\ref{eq:SNeq}), we estimate numerical values of the SN solutions $X^{in}_{lm\hat{\omega}}$ and $X^{up}_{lm\hat{\omega}}$. For greater detail, readers are also suggested to refer \cite{10.1093/ptep/ptaa149, PhysRevD.102.024041}.

\subsection{Orbit-evolution and gravitational wave phase}\label{teu2n}

As we have already set up the required ingredients for examining the orbital evolution and generated gravitational wave phase, here, we provide the final expressions for total energy flux and phase for the orbital evolution of the secondary in the EMRI system. We also consider the adiabatic evolution of the fluxes due to the gravitational back-reaction effect. The total (normalized) flux is written as,
\begin{align}\label{flux_formula}
    \mathcal{F} = \frac{1}{q}\Bigg[\left(\frac{d\hat{E}}{d\hat{t}}\right)^{H}_{\textrm{GW}} + \left(\frac{d\hat{E}}{d\hat{t}}\right)^{\infty}_{\textrm{GW}}\Bigg] 
\end{align}
which is independent of mass-ratio in the leading order of mass-ratio \cite{PhysRevD.102.024041}. Further, the evolution of the orbit takes the following form,
\begin{align}
    \frac{d\hat{r}}{d\hat{t}} &= -q \mathcal{F}(\hat{r})\Big(\frac{d\hat{E}}{d\hat{t}}\Big)^{-1}\label{orb1}\\
    \frac{d\phi}{d\hat{t}} &= \hat{\Omega}(\hat{r}(\hat{t}))\label{orb2}.
\end{align}
The solution of the adiabatic evolution of the orbit Eq.~(\ref{orb1}) provides the instantaneous
orbital phase, which is directly connected to the dominant mode GW phase by $\Phi_{GW}(\hat{t})=2\phi(\hat{t})$. Taking the initial point of the start $\hat{r}^{initial}$, we consider the evolution domain $\hat{r}\in (\hat{r}^{initial},\hat{r}^{ISCO})$. We also note that when the object crosses the ISCO, the adiabatic approximation is no longer valid. Next, we provide the results for flux and phase computations.

\section{Numerical method and Results}\label{rslt}
In this section, we summarize our method to obtain the gravitational waveform. As described in Section (\ref{apenteu3}), we obtain $R^{\textup{in},\textup{up}}_{\ell m \hat{\omega}}(\hat{r})$ by integrating Eq.~(\ref{eq:SNeq}) with boundary condition mentioned in Eq.~(\ref{eq:inBCSN}) and Eq.~(\ref{eq:upBCSN}) and using the relation mentioned in Eq.~(\ref{rinsol}). The spheroidal harmonics $S_{lm\hat{\omega}}(\theta)$ and the eigenvalue of the angular Teukolsky equation $ \lambda_{lm\hat{\omega}}$ is obtained by using \texttt{Black hole Perturbation Toolkit} package \cite{BHPT}. The numerical evaluation of the energy flux mentioned in Eq.~(\ref{flux_inf}) and Eq.~(\ref{flux_hor}) requires us to truncate the infinite sum presented in this equation. 
In this paper, we truncate the infinite sum at $l_{\textrm{max}}=10$. 
For each value of $l$, we include all $m$ values from $1$ to $l$. The total flux is obtained using the relation mentioned Eq.~(\ref{flux_formula}). In Table~(\ref{tab:truncation}), we report the truncation error $\Delta \mathcal{F}^{\textrm{tr}}=\bigg|\mathcal{F}^{l=11}-\mathcal{F}^{l=10}\bigg|/\mathcal{F}^{l=11}$ where $\mathcal{F}^{l=11}$ and $\mathcal{F}^{l=10}$ correspond to the flux calculated considering $l_{\textrm{max}}=11$ and $l_{\textrm{max}}=10$ respectively. As can be seen, the truncation error is negligible.
\begin{table}[t!]
\centering
\begin{tabular}{cc|cc|cc|cc}\hline
\hline
\multicolumn{2}{c|}{$Q=0$}                                   & \multicolumn{2}{c|}{$Q=0.01$}                               & \multicolumn{2}{c|}{$Q=0.1$}                                 & \multicolumn{2}{c}{$Q=0.5$}                                  \\ \hline
\multicolumn{1}{c}{$\hat{r}$}    & $\Delta \mathcal{F}^{\textrm{tr}}$                         & \multicolumn{1}{c}{$\hat{r}$}    & $\Delta \mathcal{F}^{\textrm{tr}}$                            & \multicolumn{1}{c}{$\hat{r}$}    & $\Delta \mathcal{F}^{\textrm{tr}}$                           & \multicolumn{1}{c}{$\hat{r}$}    & $\Delta \mathcal{F}^{\textrm{tr}}$                             \\ \hline
\multicolumn{1}{c}{12.0} &$ 1.79948\times 10^{-9} $& \multicolumn{1}{c}{12.0} &$ 1.77092\times 10^{-9} $& \multicolumn{1}{c}{12.3} &$ 1.54009\times 10^{-9} $& \multicolumn{1}{c}{13.4} &$ 8.94344\times 10^{-10}$ \\
\multicolumn{1}{c}{10.}  &$ 1.02652\times 10^{-8} $& \multicolumn{1}{c}{10.0} &$ 1.00667\times 10^{-8} $& \multicolumn{1}{c}{10.3} &$ 8.49158\times 10^{-9} $& \multicolumn{1}{c}{11.4} &$ 4.40055\times 10^{-9}$  \\ 
\multicolumn{1}{c}{8.00} &$ 8.97426\times 10^{-8} $& \multicolumn{1}{c}{8.03} &$ 8.74973\times 10^{-8} $& \multicolumn{1}{c}{8.29} &$ 7.02143\times 10^{-8} $& \multicolumn{1}{c}{9.39} &$ 3.03693\times 10^{-8} $ \\
\multicolumn{1}{c}{6.00} &$ 1.67739\times 10^{-6} $& \multicolumn{1}{c}{6.03} &$ 1.61629\times 10^{-6} $& \multicolumn{1}{c}{6.29} &$ 1.17423\times 10^{-6} $& \multicolumn{1}{c}{7.39} &$ 3.61907\times 10^{-7} $\\ \hline
\bottomrule
\end{tabular}
\caption{The fractional truncation error $\Delta \mathcal{F}^{tr}$ with $q=3\times 10^{-5}$ by comparing the fluxes truncated at $l_{max}$=10 with the ones truncated at $l_{max}$=11.} \label{tab:truncation}
\end{table}

 In order to check the accuracy of our numerical code, we compare our results for gravitational wave flux for \Schld\ case $Q=0$ with those available in \cite{BHPT, PhysRevD.90.084025}, which agrees with our results quite well.  Further, in Appendix~(\ref{t2n}), in  we provide a Table ~(\ref{t2}) to establish a comparison with \cite{BHPT, PhysRevD.90.084025}. 
 
 \begin{figure}[b!]
		\centering
		\minipage{0.5\textwidth}
		\includegraphics[width=\linewidth]{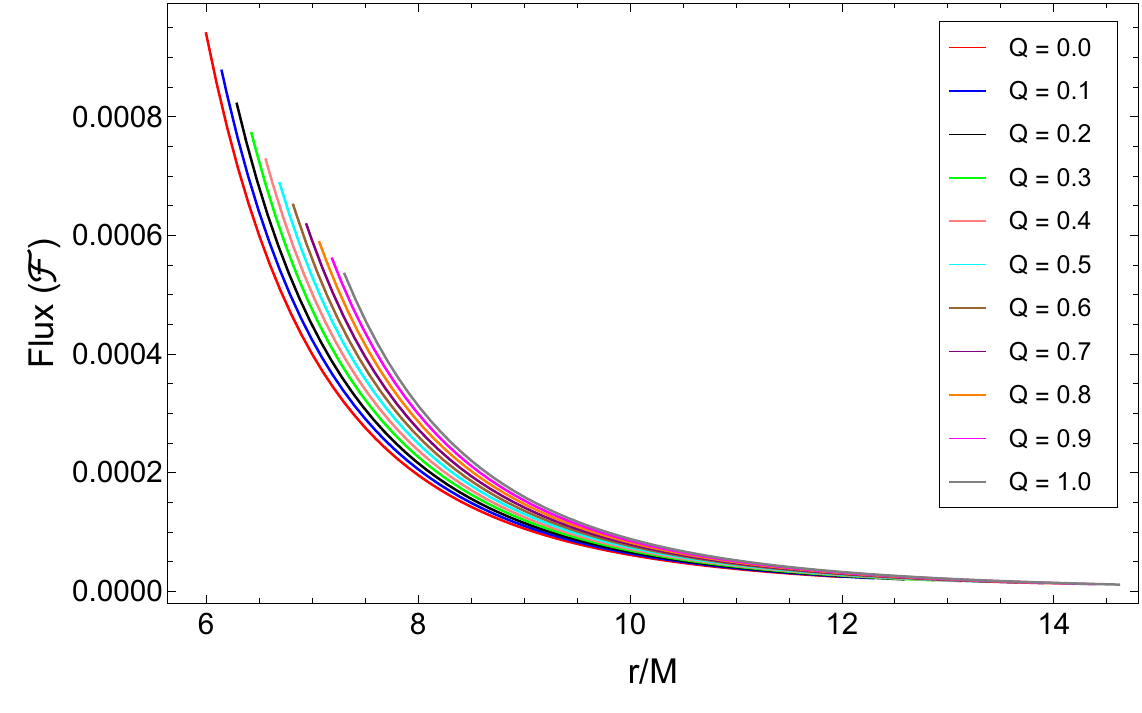}
		\endminipage\hfill
		\minipage{0.5\textwidth}%
		\includegraphics[width=8.5cm, height=5.1cm]{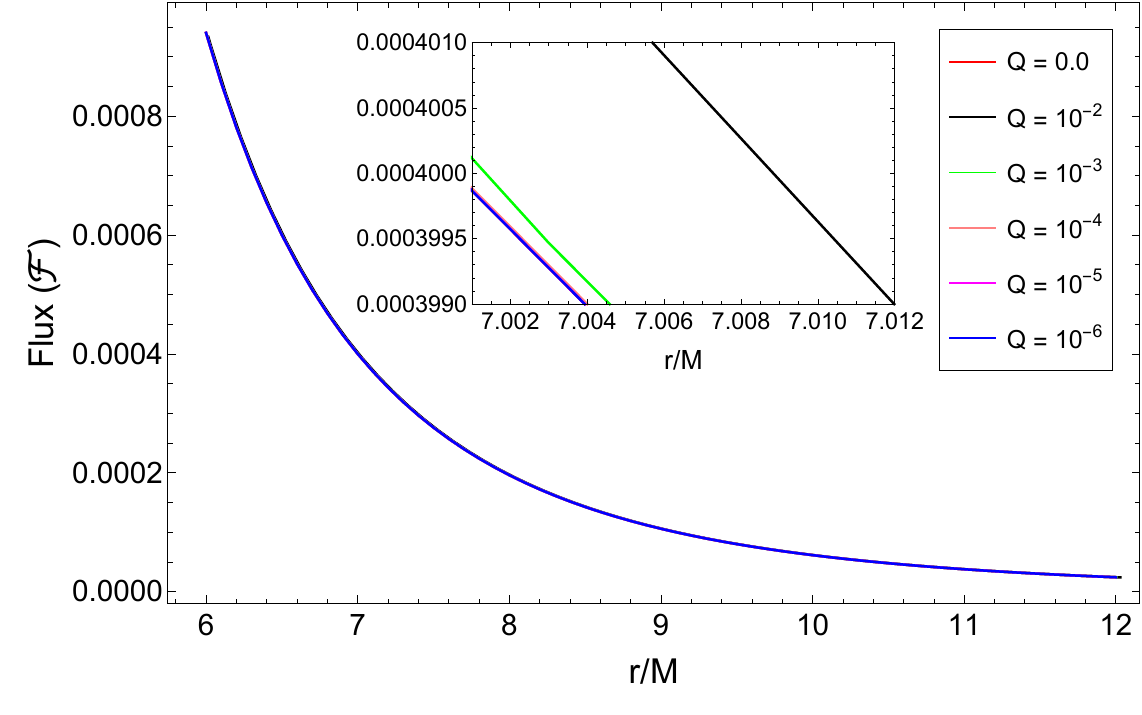}
		\endminipage
		\caption{Plots of total energy flux ($\mathcal{F}$) as a function of orbital radius ($\hat{r}$) for distinct values of $Q$.} \label{fig:flux}
\end{figure}
 
 \subsection{Gravitational wave flux and orbital phase}   
 
 In Fig.~(\ref{fig:flux}), we present the plot of gravitational flux as a function of orbital radius $\hat{r}$ for two sets of tidal charge parameter values: one for relatively large values of tidal charge parameter $Q\in \left[0.1,1\right]$ (left plot) and one for smaller values of tidal charge $Q\in \left[10^{-6},10^{-2}\right]$ (right plot). In both cases, we compare the flux value with that of a Schwarzschild black hole, corresponding to $Q=0$. Throughout the paper, we assume the inspiral of $30~M_\odot$ black hole into a supermassive black hole of mass $10^{6}~M_\odot$, such that the mass-ratio is $q=3\times 10^{-5}$. However, we note that the normalized gravitational wave flux is independent of the mass-ratio in the leading order of $q$ (see \ref{flux_formula}).
\begin{figure}[t!]
		\centering
		\minipage{0.50\textwidth}
		\includegraphics[width=8.5cm, height=5.1cm]{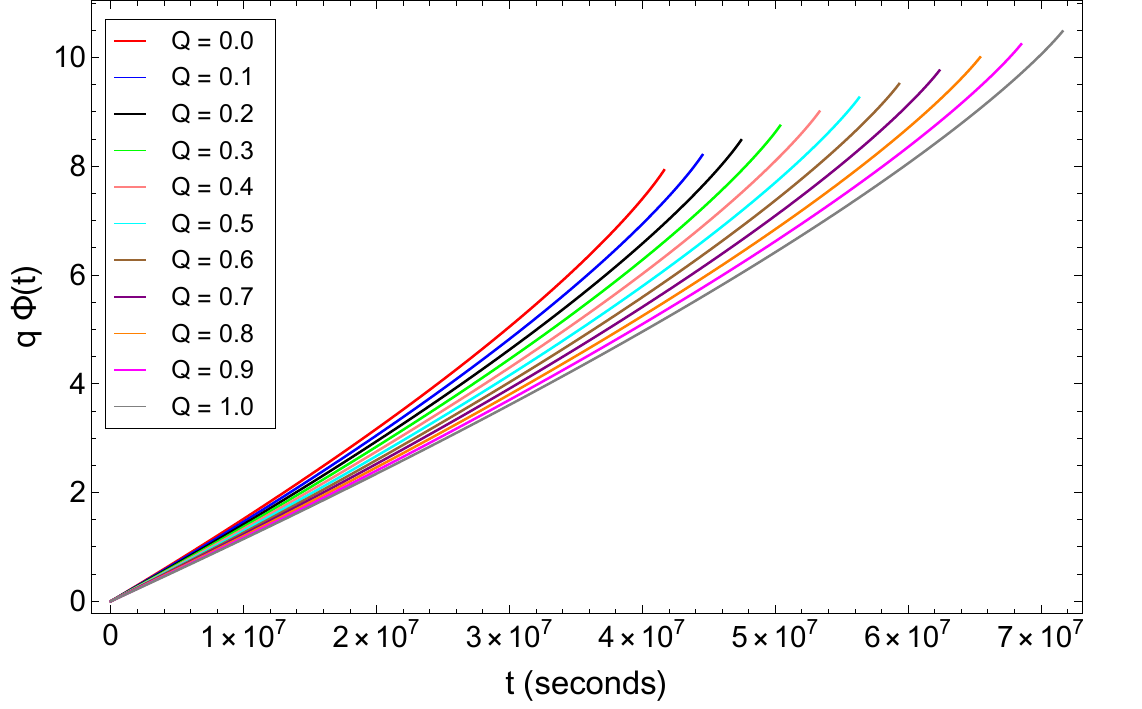}
		\endminipage\hfill
		\minipage{0.50\textwidth}%
		\includegraphics[width=8.5cm, height=5.1cm]{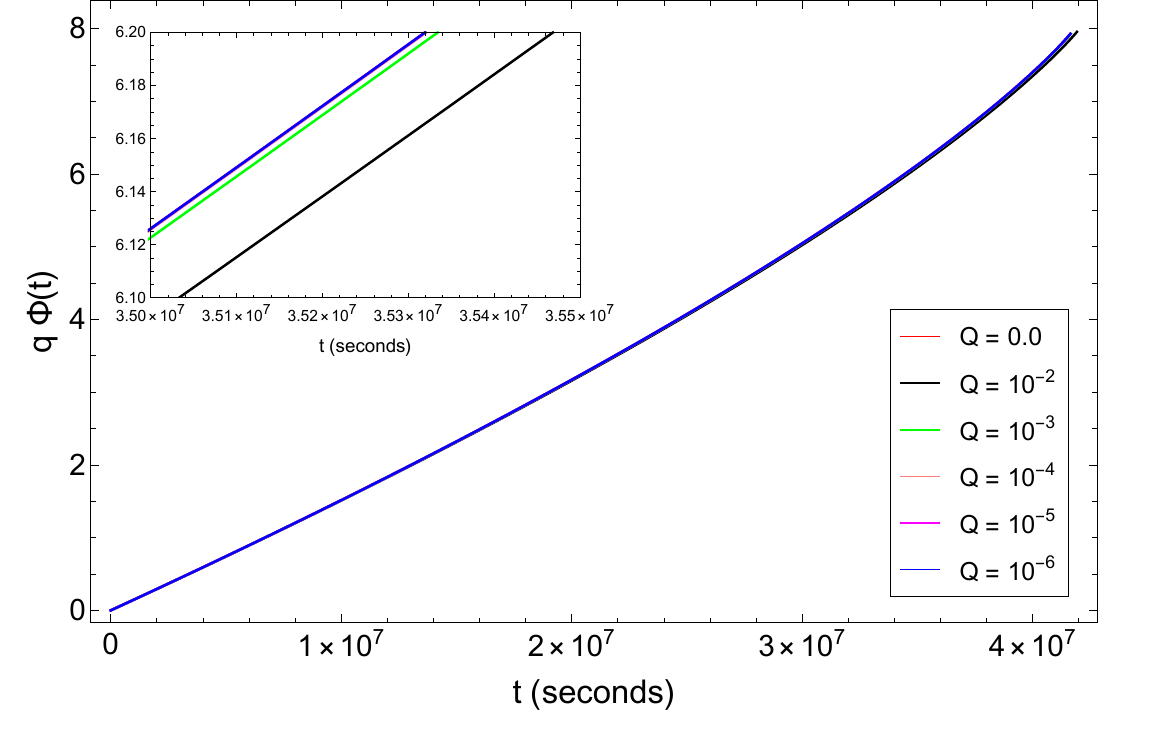}
		\endminipage
		\caption{Plots for the change of inspiralling phase of the secondary over time for different values of $Q$.} \label{fig:phase}
\end{figure}

\par
\begin{figure*}[htb!]
\centering
	\minipage{0.5\textwidth}%
	\includegraphics[width=8.2cm, height=5.15cm]{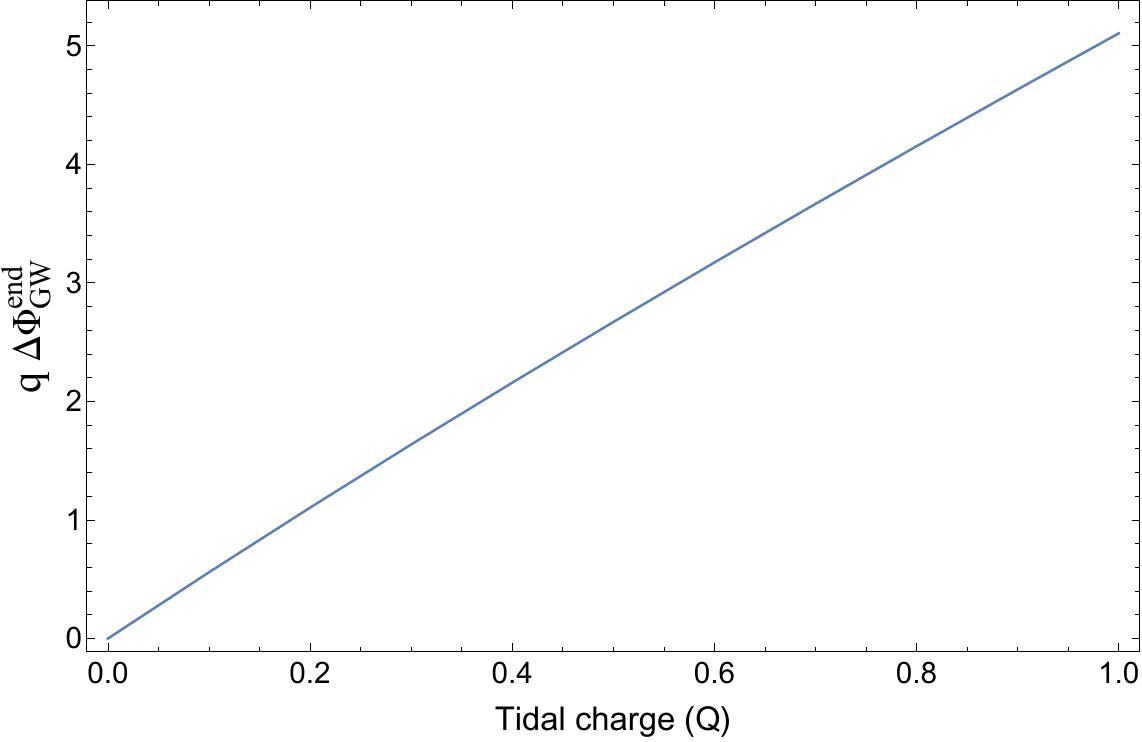}
	\endminipage
	\caption{Change in GW phase ($\Delta\Phi^{\text{end}}_{\text{GW}}$) at the ISCO for different values of the tidal charge $Q$.}\label{QNMvsP0n}
	\end{figure*}

 The evolution of orbital radius is obtained by numerically integrating Eq.~(\ref{orb1}) in the range $\hat{r}\in \left(\hat{r}^{\textrm{initial}},\hat{r}^{ISCO}\right)$ with  $\hat{r}^{\textrm{initial}}=2\hat{r}^{ISCO}$. The inspiral ends when the secondary object reaches the ISCO radius. We obtain the evolution of the orbital phase by replacing the value of $\hat{r}(\hat{t})$ in Eq.~(\ref{orb2}) and integrating  it with initial condition $\phi(t=0)=0$. Note that, in the leading order in mass-ratio $q$, we can write the \textit{instantaneous} orbital phase as $\phi(\hat{t})=\phi_0(\hat t)/q$, where the $\phi_0$ is a numerical parameter\cite{PhysRevD.78.064028}. In Fig.~(\ref{fig:phase}), we show the evolution of the $\phi_0$ for different values of $Q$. As can be seen, the inspiral time increases with the increase of the tidal charge parameter. The instantaneous orbital phase is related to the dominant mode GW phase through the relation $\Phi_{GW}(\hat{t})=2\phi(\hat{t})$. We denote the \textit{accumulated} GW phase at the end inspiral by $\Phi_{GW}^{\textit{end}}\equiv \Phi_{GW}^{0,\textit{end}}/q=\Phi_{GW}(\hat{t}_{\textrm{end}})$, which corresponds to the GW phase at $\hr=\hat{r}^{ISCO}$. Here, $\hat{t}_{\textrm{end}}$ represents the time at which the secondary object reaches $ \hat{r}^{ISCO}$. Note that the numerical parameter $\Phi_{GW}^{0,\textit{end}}$ depends only on the tidal charge parameter.
In Fig.~(\ref{QNMvsP0n}), we show that GW phase shift due to the presence of tidal charge $\Delta \Phi_{GW}^{0,\textit{end}}=q\Delta \Phi_{GW}^{\textit{end}}=\Phi_{GW}^{0,\textit{end}}(Q)-\Phi_{GW}^{0,\textit{end}}(Q=0)$ as a function of $Q$. We fit the numerical data for $\Delta \Phi_{GW}^{0,\textit{end}}(Q)$ with a cubic polynomial 
\begin{equation}
  \Delta\Phi_{GW}^{0,\textit{end}}(Q)=\sum_{i=1}^{3} a_i Q^i
\end{equation}
 with $a_1=5.65147$, $a_2=-0.699357$, and $a_3=0.152862$. This fit is accurate within  $0.12\%$ in the range $Q\in [0,1]$.


\begin{figure*}[tbh!]
	\centering
	\minipage{0.348\textwidth}
	\includegraphics[width=\linewidth]{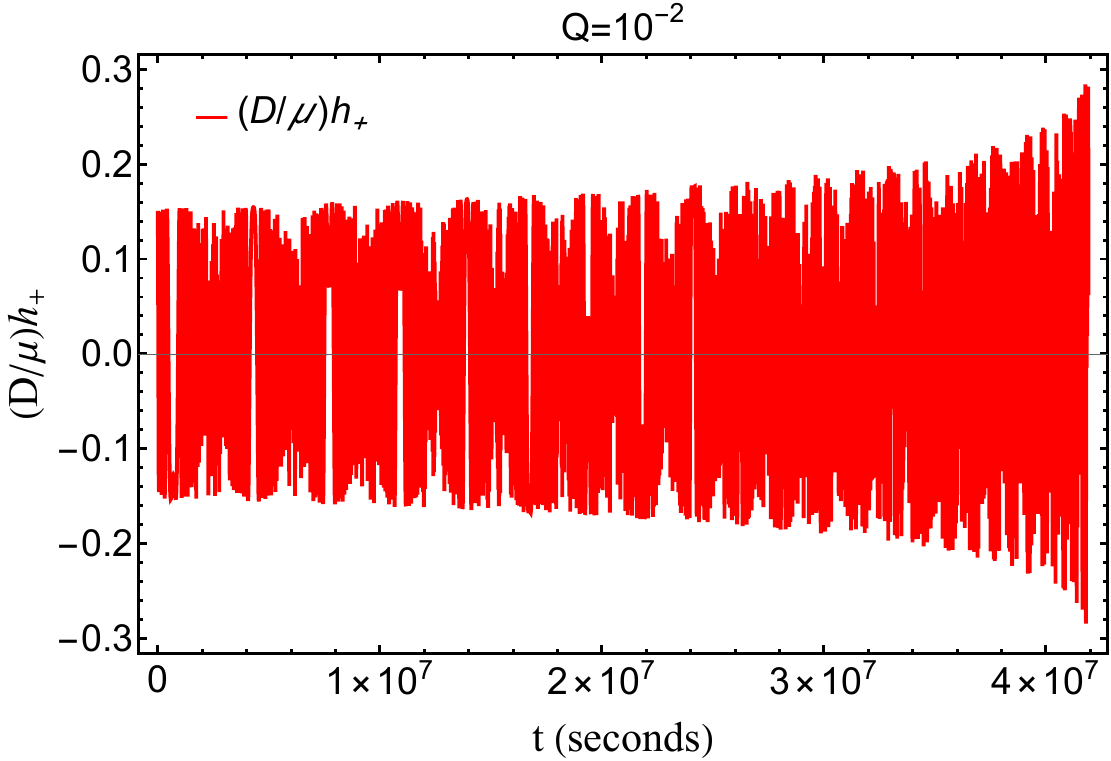}
	\endminipage
	\minipage{0.33\textwidth}%
	\includegraphics[width=\linewidth]{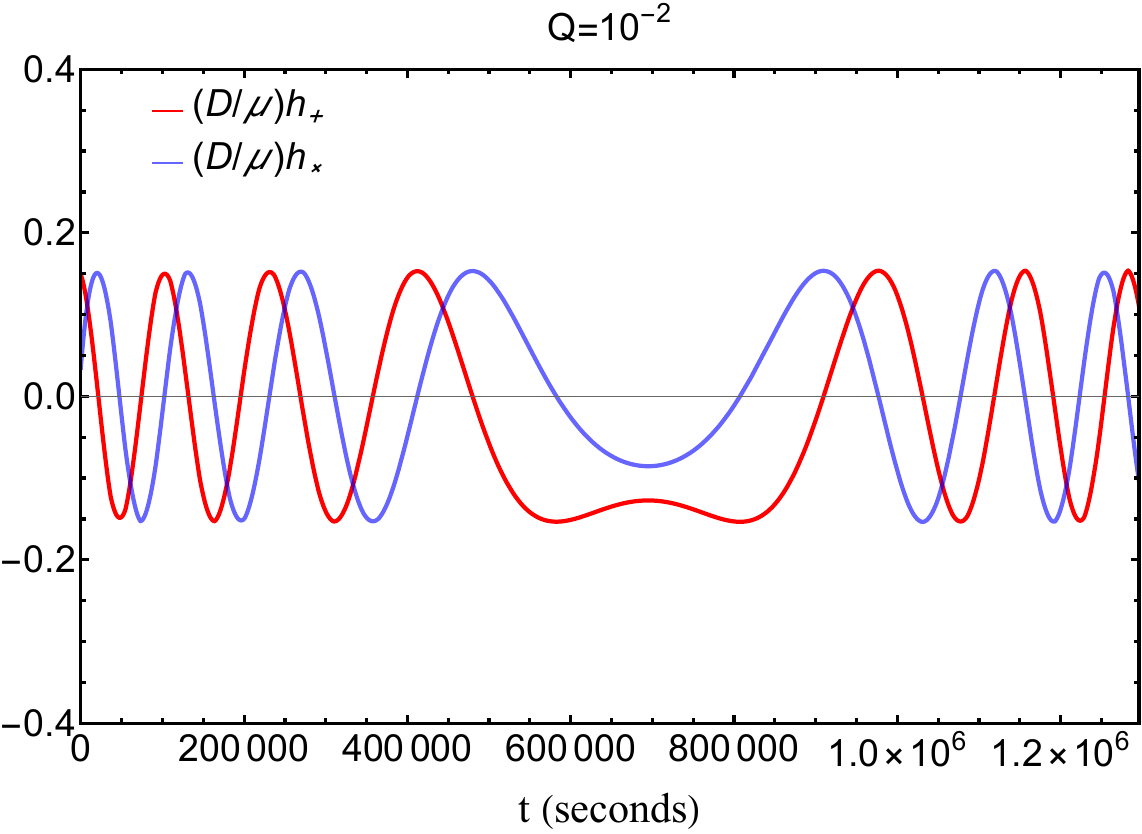}
	\endminipage
	\minipage{0.33\textwidth}%
	\includegraphics[width=\linewidth]{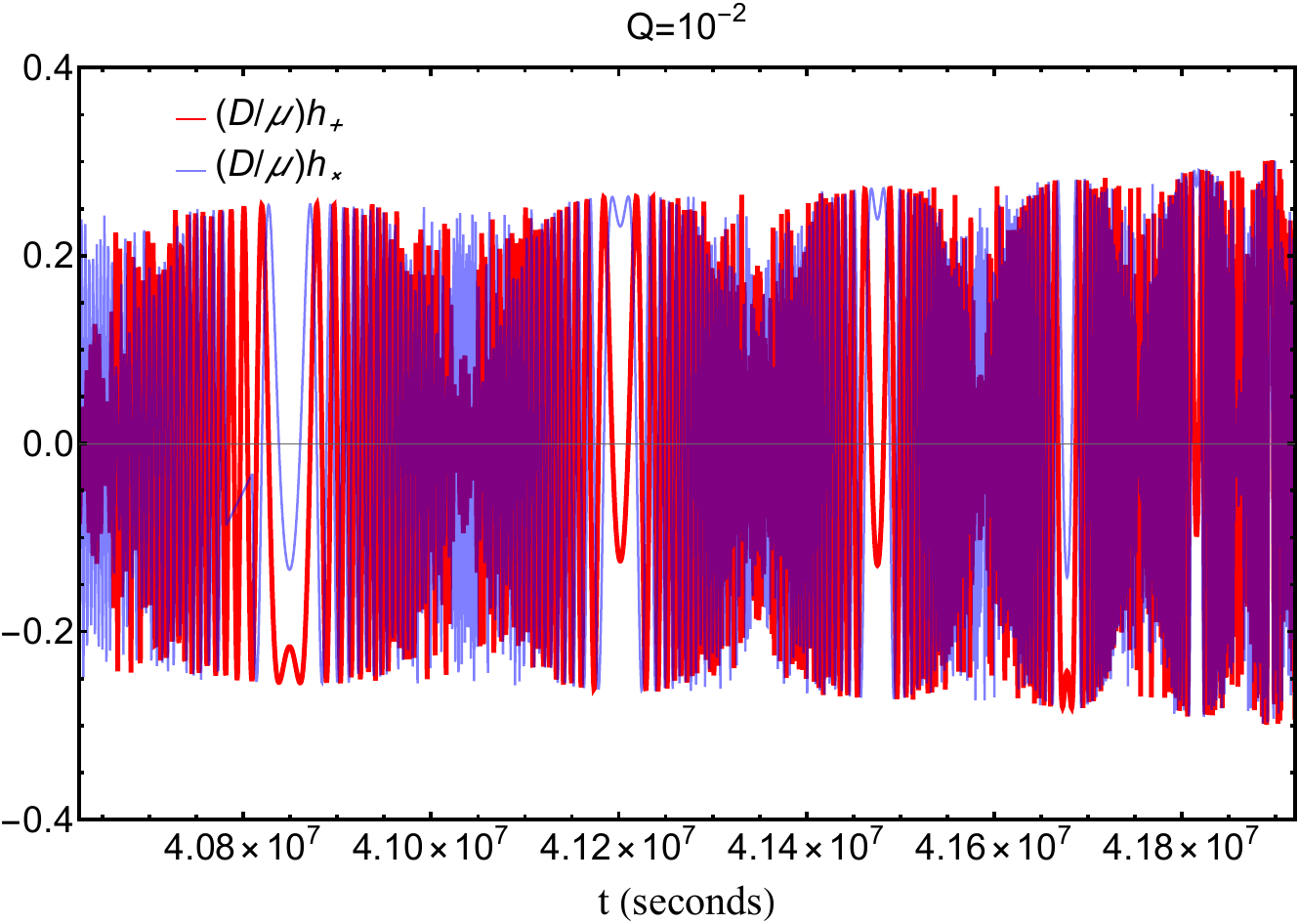}
	\endminipage\hfill
	\minipage{0.348\textwidth}
	\includegraphics[width=\linewidth]{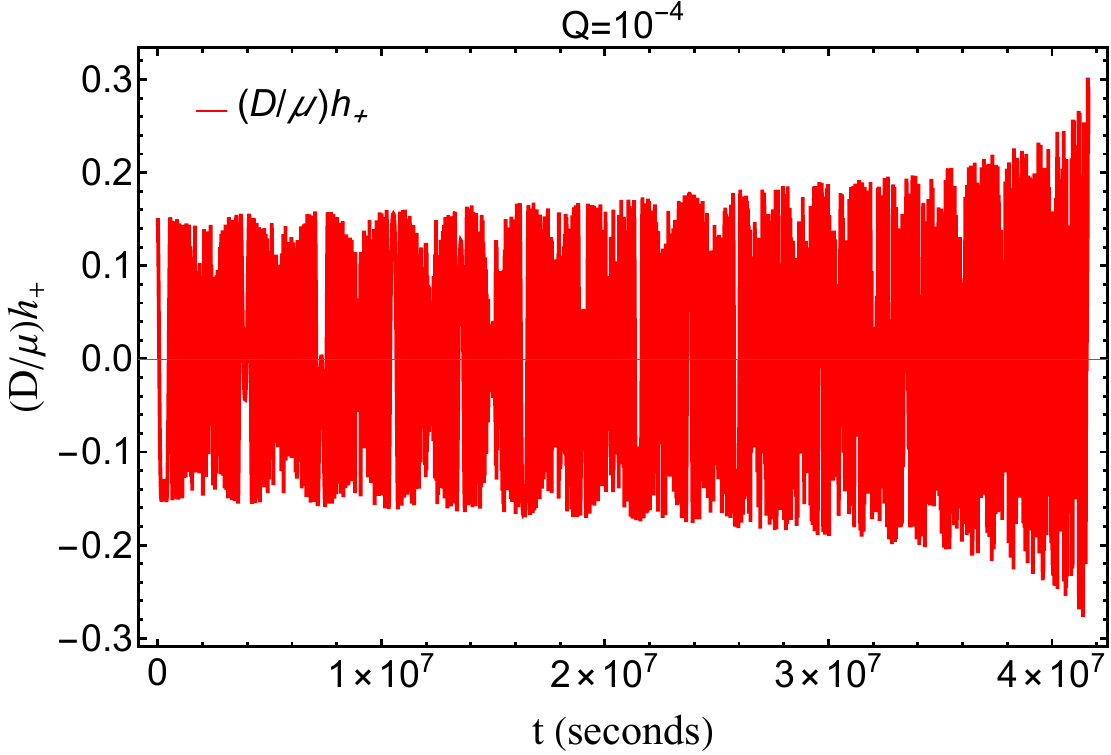}
	\endminipage
	\minipage{0.33\textwidth}%
	\includegraphics[width=\linewidth]{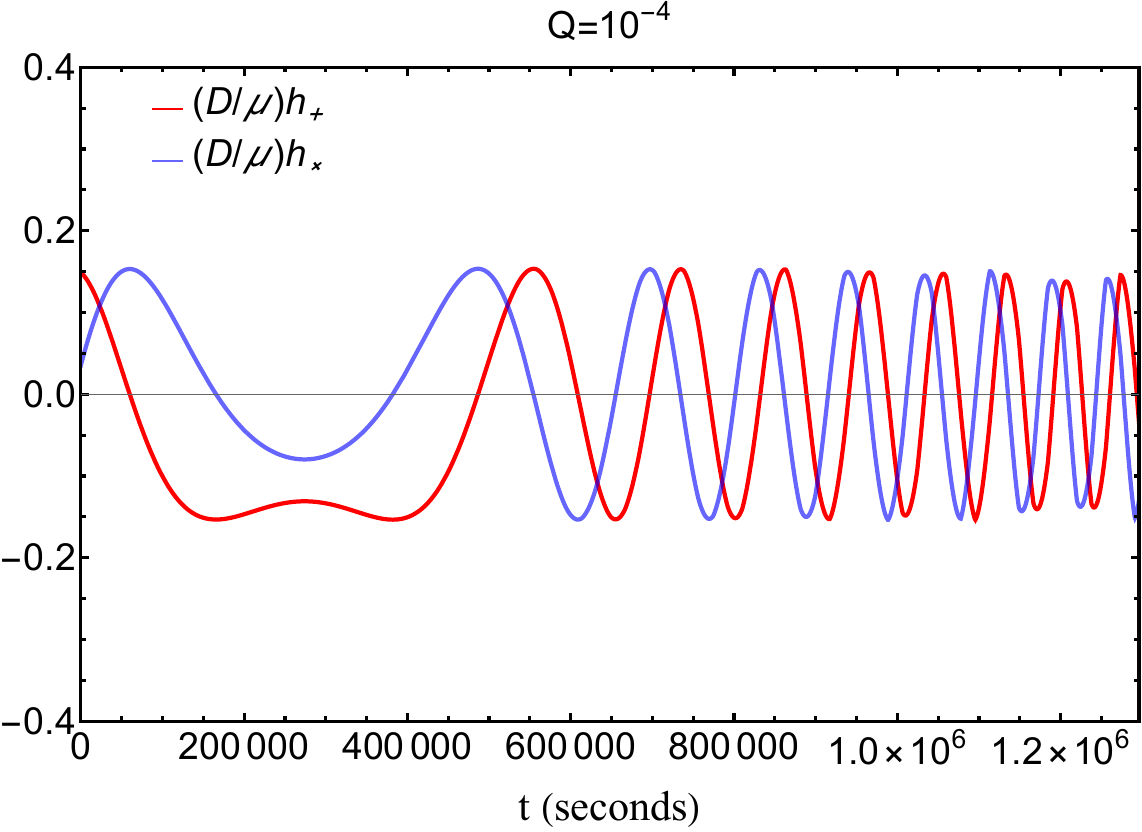}
	\endminipage
	\minipage{0.33\textwidth}%
	\includegraphics[width=\linewidth]{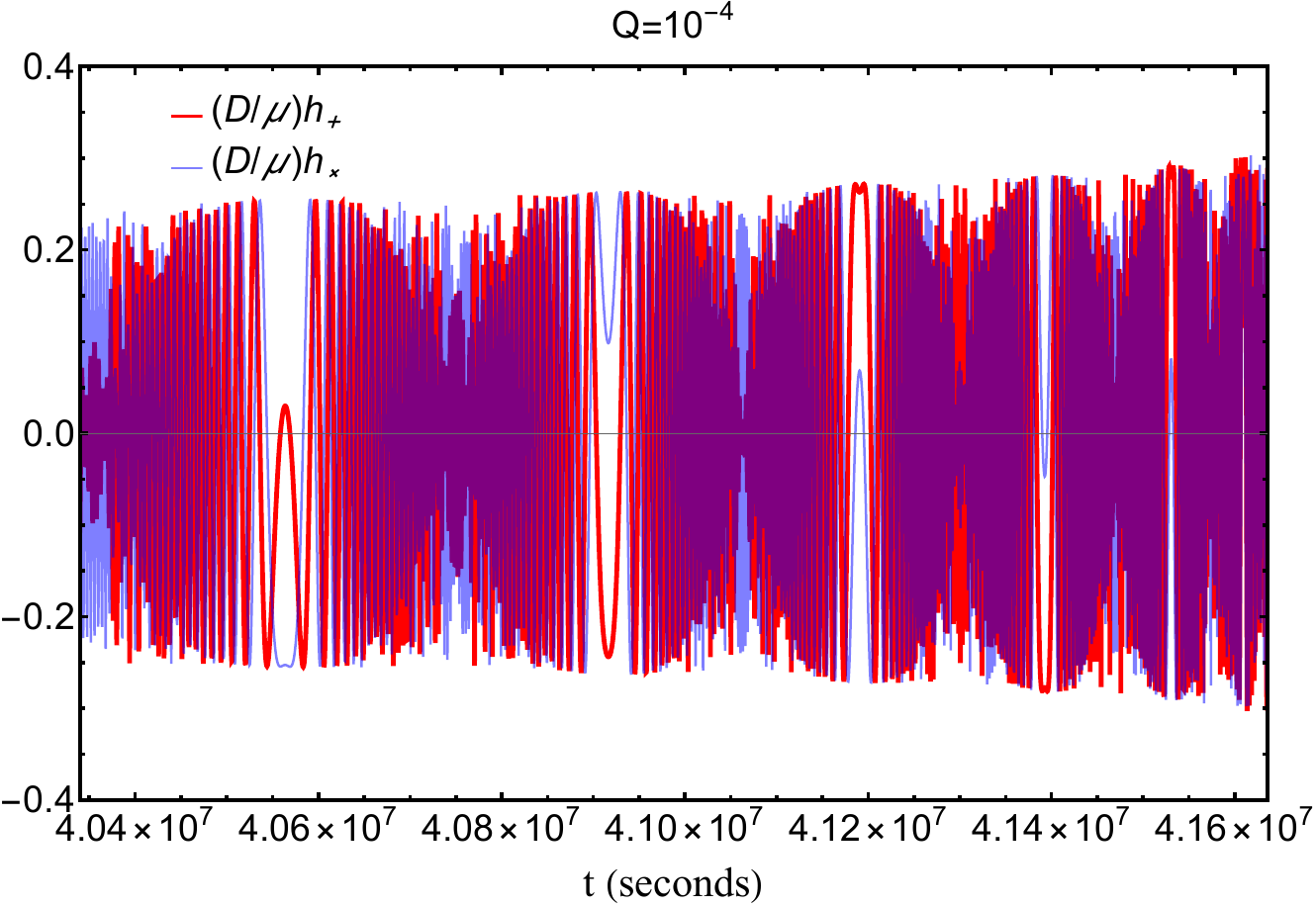}
	\endminipage
	\caption{Gravitational wave signal from the inspiral of a $30~M_\odot$ compact object into a supermassive black hole of mass $10^6~M_\odot$ for $Q=10^{-2}$ (top panel) and $Q=10^{-4}$ (botom panel). We consider the inspiral starts at $2r^{\textrm{ISCO}}$ and ends when the compact object reaches $r^{\textrm{ISCO}}$ (beyond this point, adiabatic approximation breaks down). In each panel, the leftmost plot depicts the waveform $(D/\mu)h_+$ over the whole inspiral period, where $D$ is the luminosity distance from the source to the detector, and  $\mu$ is the mass of the compact object. The plots in the middle represent the waveform for the first 15 days since the start of the inspiral, whereas the rightmost plots depict the same for the last 15 days of the inspiral.}\label{fig:Signal_Q00n}
	\end{figure*}
\subsection{Waveforms}
The emitted gravitational waveform has the following form \cite{Piovano:2021iwv, Gourgoulhon:2019iyu} 
\begin{align}\label{signal_observer}
		h(t)\equiv h_+-i h_\times(t)=-\frac{2\mu}{D}\sum_{l=2}^{\infty}\sum_{m=-l}^{l}A_{lm\Omega}(t)~S_{lm{\Omega}}(\vartheta) e^{-im\Phi(t)}.
\end{align}
where $A_{lm\Omega}(t)\equiv \mathcal A^{H}_{lm{\Omega}(t)}/(m{\Omega(t)})^2$, $\Phi(t)=\phi(t)- (\varphi+\phi_i)$, $(\vartheta, \varphi)$ is the direction of the detector in the  Boyer-Lindquist coordinates, and $D$ is the luminosity distance from source to detector. Since, the initial phase $\phi_i=\phi(t=0)$ is degenerate with azimuthal direction $\varphi$, we set $\phi_i=\varphi=0$. At a large distance from the source $D\gg M$, the dominant contribution to the gravitational waveform comes from the $(l,m)=(2,\pm 2)$ modes \cite{Gourgoulhon:2019iyu}. Thus, keeping the contribution of $(l,m)=(2,\pm 2)$ modes in the summation of Eq.~(\ref{signal_observer}), we find 
\begin{equation}\label{signal_observer_22}
 \begin{aligned}
		 h(t)=B_{22\Omega}^+\left[S_{22\Omega}(\vartheta)+S_{22\Omega}(\pi-\vartheta)\right]-i B_{22\Omega}^\times\left[S_{22\Omega}(\vartheta)-S_{22\Omega}(\pi-\vartheta)\right]
\end{aligned}   
\end{equation}
where, $B_{22\Omega}^+=|A_{lm\Omega}|\cos{(m\phi-\psi)}$ and $B_{22\Omega}^\times=|A_{lm\Omega}|\sin{(m\phi-\psi)}$ with $|A_{lm\Omega}|$ and $\psi$ being the absolute value and argument of  $A_{lm\Omega}$. Here, we use the following relations: $A_{l-m-\omega}=(-1)^l~A_{lm\omega}^{*}$ and $S_{l-m-\omega}(\vartheta)=(-1)^l~S_{lm\omega}(\pi-\vartheta)$. In Fig.~(\ref{fig:Signal_Q00n}), we present the gravitational wave signal for different values of tidal charge. Here also, we consider the inspiral of a $30~M_\odot$ black hole into $10^6~M_\odot$ supermassive black hole. 




\begin{figure*}[t!]
	\centering
	\minipage{0.60\textwidth}
	\includegraphics[width=\linewidth]{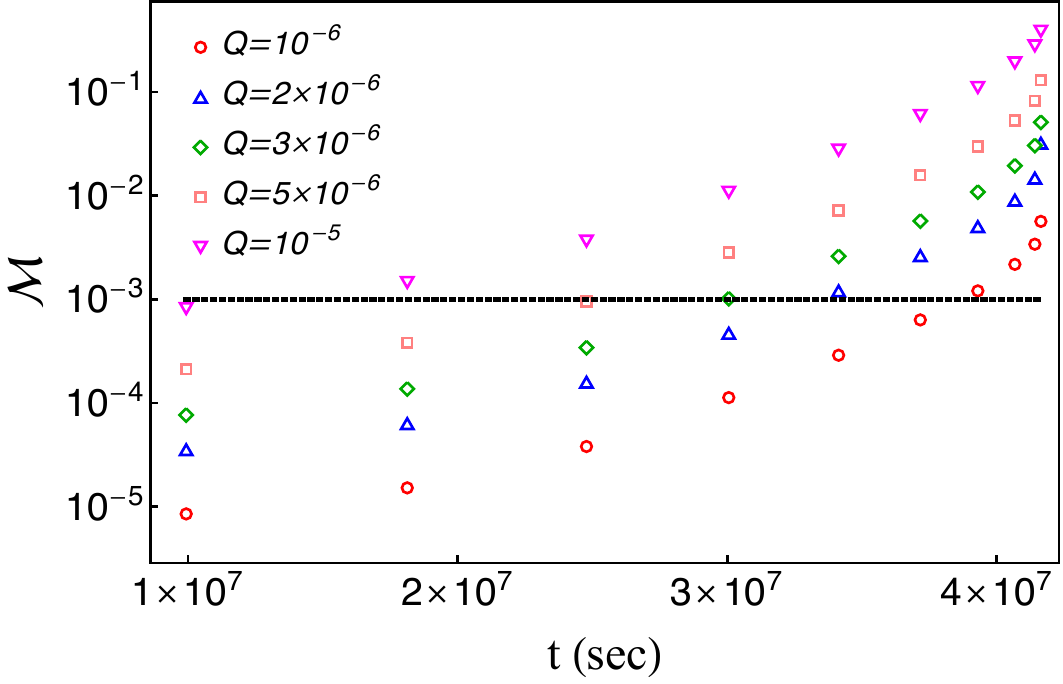}
	\endminipage
	\caption{The mismatch between the gravitational waveform originating from an EMRI system with primary object described by $M=10^{6}M_{\odot}$ and secondary with $\mu=30 M_{\odot}$. }\label{fig:mismatch}
	\end{figure*}

\subsection{Overlap}

Finally, we turn our discussion to overlap, describing the effects of tidal charge and the mismatch between two waveforms.  
The \textit{overlap} can be defined in the following way considering two model waveforms $h_{1}(t)$ and $h_{2}(t)$ as 
\cite{PhysRevD.78.124020},
\begin{align}\label{ovp1}
    F(h_{1}, h_{2}) = \frac{(h_{1}|h_{2})}{\sqrt{(h_{1}|h_{1})(h_{2}|h_{2})}},
\end{align}
where,
\begin{align}\label{ovp2}
(h_{1}|h_{2}) = 4 Re \Large\int_{0}^{\infty} \frac{\tilde{h}_{1}(f) \tilde{h}^{*}_{2}(f)}{S_{n}(f)} ,
\end{align} 
is the noise-weighted inner product. Further, $\tilde{h}(f)$ is the Fourier transformation of $h(t)$ and $S_{n}(f)$ is the power spectral density. The perfect agreement between waveforms are met if the overlap is $F=1$. The \textit{mismatch} can be defined as $\mathcal{M} =1 - F$. In this paper, we present an order-of-magnitude estimation of whether we can probe the tidal charge through LISA observation. Here the rule of thumb is that the waveform $h_1$ and $h_2$ will be indistinguishable through gravitational wave observation if $\mathcal{M}(h_1,h_2)<\mathcal{M}_{\textrm{crit}}\approx 1/2\rho^2$, where $\rho$ is the signal-to-noise ratio (SNR) and $\mathcal{M}_{\textrm{crit}}$ is the detection threshold \cite{Lindblom:2008cm, Flanagan:1997kp, PhysRevLett.123.101103, Maggio:2021uge}. Considering the average SNR for LISA observation $\sim 30$ \cite{Babak:2017tow}, we take the detection threshold as $\mathcal{M}_{\textrm{crit}}\approx 0.001$. Since we are interested in seeing the deviation from general relativity, we consider the waveform from an EMRI system with the primary described by a \Schld\ black hole as our standard waveform and calculate the mismatch with respect to it. In Fig.~(\ref{fig:mismatch}), we consider various values of tidal charge ($Q$) and plot the mismatch from the standard waveform. As before, we set $q=3\times 10^{-5}$. Note that, even for very small values of tidal charge $\mathcal{O}(10^{-6})$, the mismatch exceeds the threshold value well before the end of the inspiral phase. This implies EMRI observation can put much tighter constraints on the tidal charge parameter than black hole shadow observation or the gravitational wave from the coalescence of comparable mass binaries. In particular, \cite{Mishra:2021waw} put an upper bound on the tidal charge parameter $Q\lesssim 0.05$ through gravitational wave observation, while \cite{Neves:2020doc} found the bound as $Q\lesssim 0.004$ through shadow observations. Thus, LISA is much better suited for probing the existence of higher dimensions.

\section{Discussion}\label{dscn}
EMRI is a system that has gained noticeable attention with the potential detectability and impact of these sources on future detectors. On the other hand, the existence of an additional spatial dimension has distinctive characteristics on the four-dimensional brane. In this article, we showcase a detailed study of such a system that involves a spherically symmetric static brane black hole (primary) carrying a parameter called tidal charge (Q) and present the observational implications of the same through upcoming gravitational wave detectors. With this inspiration, we compute the imprints of tidal charge on gravitational wave flux, adiabatic evolution of the secondary and gravitational waveforms.\par
We begin with the study of the equatorial orbital motion and find the position of ISCO of the secondary. We also note that the adiabatic approximation will break down as soon as the secondary object crosses the ISCO region. We implement Teukolsky formalism to construct the solution of the radial perturbation equation provided the energy-momentum tensor of the non-spinning secondary. As a result, we determine the effect of tidal charge on energy flux and find appreciable differences with the Schwarzschild limit ($Q=0$) as presented in Fig.~(\ref{fig:flux}). 
We further construct the waveforms for distinct values of Q, as can be seen in Fig.~(\ref{fig:Signal_Q00n}). We calculate overlap to estimate the mismatch between two waveforms, one for Schwarzschild and another with non-vanishing Q values as shown in Fig.~(\ref{fig:mismatch}). We find a significant mismatch between the two waveforms even for very small values of tidal charge parameter $Q\sim 10^{-6}$. Considering the average SNR for LISA observation as $30$, we show that LISA can measure this level of mismatch. Till date, the tightest bound for tidal charge parameter $Q\lesssim 0.004$ comes from black hole shadow observations \cite{Neves:2020doc}. Our study shows that EMRI observation can put a much better constraint on this parameter than black hole shadow or ground-based gravitational wave observations. Thus, LISA is better suited to proving/disproving the existence of higher dimensions. Note that, in this paper, we have presented a simple order-of-magnitude estimate on the detectability of the tidal charge parameter. However, the analysis presented here ignores the possible correlation between different binary parameters. In a follow-up work, we perform Fisher-matrix analysis to have a more quantitative understanding of the measurability of the tidal charge parameter.  

The outcomes of the study implicate several new avenues to be investigated in which the extension for the rotating black holes in the brane will be an obvious step to examine the role of tidal charge and spin parameter. One can use gravitational wave observations from a single or ensemble of binaries to obtain a suitable constraint on the tidal charge and spin parameter. Furthermore, from the numerical relativity sector, one can inspect the observational checks for such black holes with Bayesian analysis for tidal charge and spin parameter. Further, in \cite{PhysRevD.102.024041}, the authors investigated the effect of spin of the secondary source on GW fluxes and GW phase, which was later extended for the spin-induced quadrupolar deformation by the authors in \cite{mosta}. 
With this motivation, another possible extension would be to analyze the effects carried out in the present article for a spinning secondary together with the inclusion of the quadrupolar deformation. We would like to pursue some of these outlooks in our upcoming studies. Such intriguing prospects will serve as prominent and challenging aspects for space-based gravitational wave detector LISA, and it might assist us in setting some strong observational constraints on available parameters of black holes.


\section*{Acknowledgements} 
The authors like to thank Takahiro Tanaka, Anjan Ananda Sen and Sumanta Chakraborty for useful discussions. The authors also like to thank the speakers of the online conference funded by Shastri Indo-Canadian Institute's Shastri Conference \& Lecture Series Grant (SCLSG) ``Testing Aspects of General Relativity," held between 11-14th March, 2022, for helpful discussion. The Research of M.R. is funded by the National Post-Doctoral Fellowship (N-PDF) from SERB, DST, Government of India (Reg. No. PDF/2021/001234). S.K. is supported by the Post-Doctoral fellowship (RES/SERB/PH/P0300/2021/0023) by Indian Institute of Technology Gandhinagar. A.B is supported by the Mathematical Research Impact Centric Support Grant (MTR/2021/000490), Start-Up Research Grant (SRG/2020/001380) by the Department of Science and Technology Science and Engineering Research Board (India) and Relevant Research Project grant (202011BRE03RP06633-BRNS) by the Board Of Research In Nuclear Sciences (BRNS), Department of atomic Energy, India.

\appendix

\section{Newmann-Penrose quantities and source term}\label{apenteu1}
Here, we provide the Newmann-Penrose quantities that are used for calculating the energy-momentum tensor as mentioned in Eq.~(\ref{source1}). The Newmann-Penrose (NP) tetrad are given as, \\
The null-tetrad can be written as,
\begin{align} \label{extrarev}
		l^{\mu}=&\left\{-\frac{\hr^2}{\hat \Delta },1,0,0\right\}
		 \hspace{3mm} ; \hspace{3mm}
		n^{\mu}=\left\{-\frac{1}{2},-\frac{\hat{\Delta }}{2 \hat{r}^2},0,0\right\}~,\\
		m^{\mu}=&\left\{0,0,\frac{1}{\sqrt{2} \hat{r}},\frac{i \csc (\theta )}{\sqrt{2} \hat{r}}\right\} \hspace{3mm} ; \hspace{3mm}
		\bar{m}^{\mu}=\left\{0,0,\frac{1}{\sqrt{2} \hat{r}},-\frac{i \csc (\theta )}{\sqrt{2} \hat{r}}\right\}~.
\end{align}
The non-vanishing spin coefficients are
\begin{equation}\label{spin_coef}
 \begin{aligned}
       \rho=\frac{1}{\hr}\,,\quad \mu=\frac{\hat \Delta}{2\hr^3}\,,\quad{\gamma}=\frac{2\hat \Delta-\hat r\hat \Delta'}{4\hr^3}\,,\quad \beta=-\alpha=-\frac{\cot\theta}{2\sqrt{2}\hr}~,
 \end{aligned}   
\end{equation}
while the non-vanishing Weyl scalar of the background spacetime is 
\begin{equation}\label{weyl_coef}
 \begin{aligned}
       \Psi_2=\frac{-12\hat \Delta+6\hat r\hat \Delta'}{12\hr^4}~.
 \end{aligned}   
\end{equation}

\subsection{Source term}\label{apenteu2}
Using Eq.~(\ref{source1}), the source term takes the following form,
\begin{equation}\label{source_term}
\mathcal {T} _ {l m \hat{\omega}} =  4 \int d\hat{t} d\theta\sin\theta d\phi \frac{\left(B' _ 2 + {B' _ 2}^*\right)}{\bar{\rho}\rho^5} S_{lm\omega} e^{-  i (m\phi+ \hat{\omega} \hat{t})}
\end{equation}
where,
\begin{align}
B' _ 2 &= - \frac{1}{2} \rho^8\bar {\rho}\mathcal {L} _{-1} 
\bigg[\frac{1}{\rho^4}\mathcal{L}_0\bigg[\frac{T_{nn}}{\rho^2\bar{\rho}} \bigg]\bigg]
 -\frac{1}{2\sqrt{2}}\hat{\Delta}^2 \rho^8\bar{\rho}\mathcal{L}_ {-1}\bigg[\frac{\bar{\rho}^2}{\rho^4} 
J_+\bigg[\frac{T_{\overline{m}n}}{ \hat{\Delta} \rho^2\bar{\rho}^2} \bigg]\bigg] \ , \\
 {B' _ 2}^*&= - \frac {1} {4}\hat{\Delta}^2 \rho^8\bar{\rho} J_+\bigg[\frac{1}{\rho^4}J_+ 
\bigg[\frac{\bar{\rho}}{\rho^2}T_{\overline{m}\overline{m}}\bigg] \bigg]
- \frac{1}{2\sqrt {2}}\hat{\Delta}^2 \rho^8\bar{\rho} J_+ \bigg[\frac{\bar{\rho}^2}{\hat{\Delta} \rho^4}\mathcal {L}_ 
{-1}\bigg[\frac{ T_ {\overline{m}n}}{\rho^2\bar {\rho}^2}\bigg] \bigg] \ ,
\end{align}
We recall that $\hat{\Delta} = \hat{r}^{2}f(\hat{r})$, $K=\hat{r}^{2}\omega$, $\rho = \bar{\rho} = \frac{1}{\hat{r}}$, and $J_+ = \frac{\partial}{\partial \hat{r}}+\frac{i K}{\hat{\Delta}}$. Further, the operators,
\begin{alignat}{2} 
\mathcal{L} _s = \frac{\partial}{\partial\theta}+\frac{m}{\sin \theta} -2 \cot 
\theta \hspace{3mm} ; \hspace{3mm} 
\mathcal{L} _s^\dagger = \frac{\partial}{\partial\theta}-\frac{m}{\sin \theta} - s 
\cot\theta.
\end{alignat}
$T_{nn}, T_ {\overline{m}n}$, and $T_{\overline{m}\overline{m}}$ are projections of the energy-momentum tensor with respect to NP tetrad, which for the point particle energy-momentum tensor defined in Eq.~(\ref{SET}) takes the values 

\begin{equation}\label{SET_NP_terms}
\begin{aligned}
    T_{nn}&\equiv T^{\mu\nu}n_{\mu}n_{\nu}=\frac{\delta^{(3)}}{\sqrt{-g}} \frac{\mu\hE}{4},\\{T_{\bm n}} &\equiv T^{\mu\nu}\bm_{\mu}n_{\nu}= \frac{\delta^{(3)}}{\sqrt{-g}}\frac{i \mu\hJ\sqrt{\hat \Delta} }{2\sqrt{2}\hr^2},\\
    {T_{\bm \bm}} &\equiv T^{\mu\nu}\bm_{\mu}\bm_{\nu}=-\frac{\delta^{(3)}}{\sqrt{-g}} \frac{ \mu\hJ^2\hat \Delta }{2\hE\hr^4}
\end{aligned}
\end{equation}
where, $\delta^{(3)}=\delta(r-\hat r(t))\delta(\theta-\theta(t))\delta(\phi-\phi(t))$ is the three-dimensional Dirac delta function. To make the comparison with \cite{PhysRevD.102.024041, mosta} more clear, we write 
\begin{equation}\label{SET_operator}
\begin{aligned}
      T_{nn}h(x)&=\delta^{(3)}~D^{\Omega}_{nn}[N_{nn}h(x)]\\
      T_{\bm n}h(x)&=\delta^{(3)}~D^{\Omega}_{\bm n}[N_{\bm n}h(x)]\\
      T_{\bm\bm}h(x)&=\delta^{(3)}~D^{\Omega}_{\bm\bm}[N_{\bm\bm}h(x)]\\
\end{aligned}
\end{equation}
where $h(x)$ is a smooth function of the coordinate variables , $N_{nn}=\hat \Delta/(\sqrt{-g}\hr^2)$,  $N_{\bm n}=\sqrt{\hat \Delta}\rho/(\sqrt{-g})$ and $N_{\bm \bm}=\hr^2~\rho^2/(\sqrt{-g})$. Quaantities like $D^{\Omega}_{nn}$ can be found by using Eq.~(\ref{SET_NP_terms}) and Eq.~(\ref{SET_operator}). Following the same procedure presented in  \cite{PhysRevD.102.024041, mosta}, we can then write the source term Eq.~(\ref{source_term}) as
\begin{align}
	\begin{split} \label{app7}
		\mathcal {T} _ {l m \hat{\omega}} =\int d\hat t\, \Delta^2\, e^{i(\hat \omega \hat t-m\,\phi)}\Big[\delta(\hat r-\hat r(\hat t))\,J_{D}^{(0)}+\partial_{\hat r}\Big(J_{D}^{(1)}\,\delta(\hat r-\hat r(\hat t))\Big)+\partial_{\hat r}^2\Big(J_{D}^{(2)}\,\delta(\hat r-\hat r(\hat t))\Big)\Big]\Big |_{\theta=\theta(\hat t),\phi=\phi(\hat t)},
	\end{split}
\end{align}
where,
\begin{align}
	\begin{split} \label{app8}
		J_{D}^{(0)}&=D^{\Omega}_{nn} \Big(f^{(0)}_{nn}\Big)+D^{\Omega}_{\bm n} \Big(f^{(0)}_{\bm n}\Big) +D^{\Omega}_{\bm \bm} \Big(f^{(0)}_{\bm\bm}\Big),\\
		J_{D}^{(1)}&=D^{\Omega}_{\bm n} \Big(f^{(1)}_{\bm n}\Big)+D^{\Omega}_{\bm \bm} \Big(f^{(1)}_{\bm \bm}\Big) \hspace{2mm} ; \hspace{2mm} J_{D}^{(2)}=D^{\Omega}_{\bm \bm}\Big(f^{(2)}_{\bm \bm}\Big),
	\end{split}
\end{align}
and 
\begin{align}
	\begin{split}
		f^{(0)}_{nn}&=-\frac{2\,\bar\rho}{\hat{\Delta}\, \rho}\Big[\mathcal{L}^{\dagger}_{1}\mathcal{L}^{\dagger}_{2}S_{lm\hat{\omega}}\Big] \hspace{3mm} ; \hspace{3mm}
		f^{(0)}_{\bm n}=\frac{4\, \bar\rho}{\sqrt{2}\rho\sqrt{\Delta}}\Big\{\Big(i\,\frac{K}{\hat{\Delta}}+\rho+\bar \rho
		\Big)\mathcal{L}_2^{\dagger}\, S_{l m\hat{\omega}}\Big\} \\
		f^{(1)}_{\bm n}&=\frac{4\, \bar\rho}{\sqrt{2}\rho\sqrt{\hat{\Delta}}}\mathcal{L}_2^{\dagger}\,S_{l m\hat{\omega}} \hspace{3mm} ; \hspace{3mm}
		f^{(0)}_{\bm \bm}=\frac{\bar \rho}{\rho}\Big[\frac{d}{d \hat{r}}\Big(\frac{i\, K}{\hat{\Delta}}\Big)-2\rho\frac{i\, K}{\hat{\Delta}}+\frac{K^2}{\hat{\Delta}^2}\Big]S_{l m\hat{\omega}},\\ f^{(1)}_{\bm \bm}&=-\Big(\bar \rho+\frac{\bar \rho}{\rho}\frac{i\, K}{\hat{\Delta}}\Big)S_{l m\hat{\omega}},\quad{f}^{(2)}_{\bm \bm}=-\frac{\bar \rho}{\rho}S_{lm\hat{\omega}}.
	\end{split}
\end{align}
With the source term for Teukolsky equation, we can write the amplitude given in Eq.~(\ref{amp_def}),
\begin{align}
	\begin{split}
		\mathcal{Z}^{H,\infty}_{l m \hat{\omega}}
		=& C^{H,\infty}_{l m \hat{\omega}}\int_{\hat{r}_+}^{\infty}  d\hat{r}'\int_{-\infty}^{\infty} d\hat{t}\, e^{i(\hat \omega \hat t-m\,\phi)}\Big[\delta(\hat r-\hat r(\hat t))\,J_{D}^{(0)}+\partial_{\hat r}\Big(J_{D}^{(1)}\,\delta(\hat r-\hat r(\hat t))\Big)\Big]R^{\textrm{in},\textrm{up}}_{l m \hat{\omega}}(\hat r')\,
	\end{split}
\end{align}
It is remind that the integration should be evaluated at $\theta=\theta(\hat t),\, \phi=\phi(\hat t)$.
Upon integration with respect to $\hat{r}$ and making use of delta function, we get 
	\begin{equation}\label{app_amplitude}
		\mathcal{Z}^{H,\infty}_{l m \hat{\omega}}
		=C^{H,\infty}_{l m \hat{\omega}}\int_{-\infty}^{\infty} d\hat{t}\, e^{i(\hat \omega \hat t-m\,\phi)}\Big[A_0-A_1\frac{d}{d\hat r}+A_2\frac{d^2}{d\hat r^2}\Big]R^{\textrm{in},\textrm{up}}_{l m \hat{\omega}}(\hat r),
	\end{equation}
	where,
	\begin{align}
		\begin{split}\label{app12}
			A_0=&O_{nn}f^{(0)}_{NN}+O_{\bm n}f^{(0)}_{\bm n}+O_{\bm \bm}f^{(0)}_{\bm \bm},\\
			A_1=&O_{\bm n}f^{(1)}_{\bm n}+O_{\bm \bm}f^{(1)}_{\bm \bm}, \quad A_2=O_{\bm \bm}f^{(2)}_{\bm \bm},
		\end{split}
	\end{align}
In the equatorial plane, the quantities like $O_{nn}$ are defined through the relation $D^{\Omega}_{nn}=O_{nn}$.

\section{Comparison of GW fluxes with previous studies} \label{t2n}

In order to verify the accuracy of our numerical code, we compare our results for gravitational wave flux for the Schwarzschild case ($Q=0$) with those available in \cite{BHPT, PhysRevD.90.084025}. In  Table (\ref{t2}), we show the comparison between our result and \cite{BHPT}.

\begin{table}[h]
\centering
\begin{tabular}{ c|c|c} 
\hline\hline
$\hat{r}$ & Flux for Schwarzschild & Flux for Schwarzschild in \cite{BHPT}\\\hline

$6$ & $9.4034 \times 10^{-4}$ & $9.3727 \times 10^{-4}$ \\  

$7$ & $4.0016 \times 10^{-4}$ & $3.9979 \times 10^{-4}$ \\

$8$ & $1.9610 \times 10^{-4}$ & $1.9604 \times 10^{-4}$ \\

$9$ & $1.0593 \times 10^{-4}$ & $1.0589 \times 10^{-4}$ \\

$10$ & $6.1516 \times 10^{-5}$ & $6.1536 \times 10^{-5}$ \\ 

$11$ & $3.7792 \times 10^{-5}$ & $3.7832 \times 10^{-5}$ \\

$12$ & $2.4292 \times 10^{-5}$ & $2.4296 \times 10^{-5}$ \\
\hline
\bottomrule
\end{tabular} \caption{\label{t2} Fluxes for the Schwarzschild case at different radial distances (with $Q = 0$) and compared to the values computed in \cite{BHPT}.}
\end{table}

\bibliography{Brane}
\bibliographystyle{JHEP}
\end{document}